\documentclass[aps,prb,preprint,onecolumn,citeautoscript]{revtex4-1} 
\usepackage{amsmath,amssymb,mathrsfs,bm,feynmf,setspace}
\usepackage{graphicx}
\usepackage[tight]{subfigure}  
\usepackage{color} 
\usepackage[papersize={8.5in,11in}]{geometry}
\usepackage[colorlinks=true]{hyperref} 
\hypersetup{
    bookmarks=true,         % show bookmarks bar?
    unicode=false,          % non-Latin characters in Acrobat's bookmarks
    pdftoolbar=true,        % show Acrobat's toolbar?
    pdfmenubar=true,        % show Acrobat's menu?
    pdffitwindow=false,     % window fit to page when opened
    pdfstartview={FitH},    % fits the width of the page to the window
    pdftitle={My title},    % title
    pdfauthor={Author},     % author
    pdfsubject={Subject},   % subject of the document
    pdfcreator={Creator},   % creator of the document
    pdfproducer={Producer}, % producer of the document
    pdfkeywords={keyword1} {key2} {key3}, % list of keywords
    pdfnewwindow=true,      % links in new window
    colorlinks=true,       % false: boxed links; true: colored links
    linkcolor=red,          % color of internal links (change box color with linkbordercolor)
    citecolor=blue,        % color of links to bibliography
    filecolor=magenta,      % color of file links
    urlcolor=cyan           % color of external links
} 
\newcommand{\al}{\alpha} 
\newcommand{\be}{\beta} 
\newcommand{\g}{\gamma}
\newcommand{\de}{\delta} 
\newcommand{\e}{\epsilon} 
\newcommand{\ve}{\varepsilon}

\newcommand{\ka}{\kappa}
\newcommand{\la}{\lambda}
\newcommand{\La}{\Lambda}

\newcommand{\s}{\sigma}

\newcommand{\w}{\omega}

\newcommand{\De}{\Delta} 
\newcommand{\G}{\Gamma}
\renewcommand{\S}{\Sigma}

\newcommand{\pd}{\partial}

\newcommand{\beq}{\begin{equation}}
\newcommand{\eeq}{\end{equation}}
\newcommand{\Beq}{\begin{eqnarray}}
\newcommand{\Eeq}{\end{eqnarray}}
\newcommand{\bml}{\begin{multline}}

\newcommand{\eeqm}{\end{multline}}

\newcommand{\bsp}{\begin{split}}
\newcommand{\esp}{\end{split}}

\newcommand{\vph}{\varphi}
\renewcommand{\b}[1]{{\bm #1}}
\renewcommand{\t}{\tilde}

\newcommand{\inv}{^{-1}}

\newcommand{\mc}{\mathcal}

\renewcommand{\t}{\tilde}
\newcommand{\ra}{\rightarrow}
\newcommand{\req}[1]{Eq.~(\ref{eq:#1})}

\newcommand{\rfig}[1]{Fig.~\ref{fig:#1}}

\DeclareMathOperator{\diag}{diag}

\newcommand{\df}{\mathsf f} %$\mathsf f \digamma \mathfrak f$
\begin{document}

\title{The quasi-normal modes of quantum criticality}
\author{William Witczak-Krempa}
\affiliation{Perimeter Institute for Theoretical Physics, Waterloo, Ontario N2L 2Y5, Canada}
\affiliation{Department of Physics, University of Toronto, Toronto, Ontario M5S 1A7, Canada}
\author{Subir Sachdev}
\affiliation{Department of Physics, Harvard University, Cambridge, Massachusetts, 02138, USA}
\date{\today\\
\vspace{1.6in}}
\begin{abstract}
We study charge transport of quantum critical points described by
conformal field theories in 2+1 spacetime dimensions. The transport is described by an 
effective field theory on an asymptotically anti-de Sitter spacetime,
expanded to fourth order in spatial and temporal gradients. The presence of a horizon at non-zero temperatures implies that this theory has
quasi-normal modes with complex frequencies.
The quasi-normal modes determine the poles and zeros of the
conductivity in the complex frequency plane, and so fully determine its behavior on the real frequency axis, at frequencies
both smaller and larger than the absolute temperature. We describe the role of particle-vortex or S-duality on the conductivity, specifically
how it maps poles to zeros and vice versa. These analyses motivate two sum rules obeyed by the quantum critical conductivity: the holographic computations are the first to satisfy {\em both\/} sum rules, while earlier Boltzmann-theory computations satisfy only one of them.
Finally, we compare our results with the analytic structure of the $O(N)$
model in the large-$N$ limit, and other CFTs.
\end{abstract}
\maketitle
\tableofcontents
\section{Introduction}

The dynamics of quantum criticality\cite{sachdev-book} has long been a central subject in the study of correlated quantum materials.
Two prominent examples of recent experiments are: ({\em i\/}) the observation of criticality in the penetration depth
of a high temperature superconductor at the quantum critical point of the onset of spin density wave order \cite{matsuda},
and ({\em ii\/}) the criticality of longitudinal ``Higgs'' excitations near the superfluid-insulator transition
of ultracold bosons in a two-dimensional lattice \cite{endres}.

A complete and intuitive description of the low temperature dynamics of {\em non\/}-critical systems is usually provided by their quasiparticle excitations.
The quasiparticles are long-lived excitations which describe all low-lying states, and their collective dynamics is efficiently captured by a quantum
Boltzmann equation (or its generalizations). The Boltzmann equation can then be used to describe a variety of equilibrium properties, such as the electrical conductivity,
thermal transport, and thermoelectric effects. Moreover, such a method can also address non-equilibrium dynamics, including the approach
to thermal equilibrium of an out-of-equilibrium initial state.

A key property of strongly-interacting quantum critical systems is the absence of well-defined quasiparticle excitations. The long lifetimes of quasiparticles is ultimately the justification of the Boltzmann equation, so a priori it appears that we cannot apply this long-established method to such quantum critical points.
However, there is a regime where, in a sense, the breakdown of quasiparticle excitations is weak: this is the limit where the anomalous exponent, usually
called $\eta$, of a particle-creation operator, $\phi$, is small (strictly speaking, $\phi$ creates particles away from the quantum critical point).
The spectral weight of the $\phi$ Green's function is a power-law continuum, but in the limit $\eta \rightarrow 0$, it reduces to a quasiparticle delta function.
By expanding away from the $\eta \rightarrow 0$ limit, one can extend to the Boltzmann method to quantum critical points, and such a method
has been the focus of numerous studies \cite{damle,ssqhe,sondhi1,sondhi2,sondhi3,sondhi4,markus1,markus2,markus3,fritz,will}.

\begin{figure}[h]
\centering%
 \subfigure[]{\label{fig:map} \includegraphics[scale=.8]{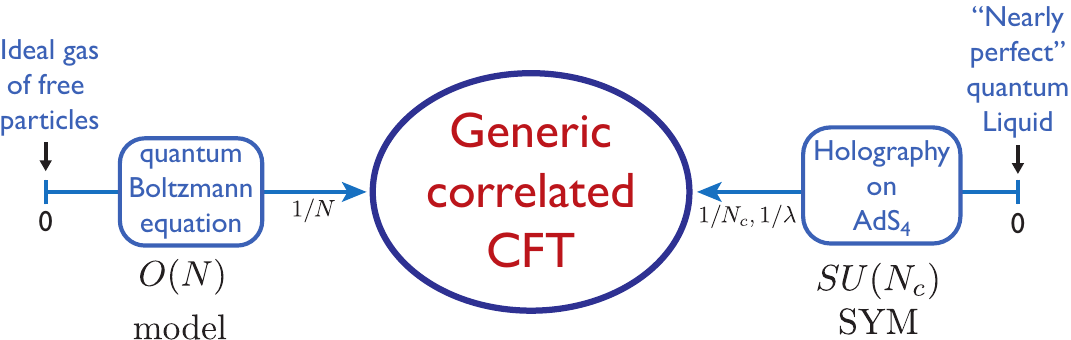}}\\
 \subfigure[]{\label{fig:sketch-rotor} \includegraphics[scale=.6]{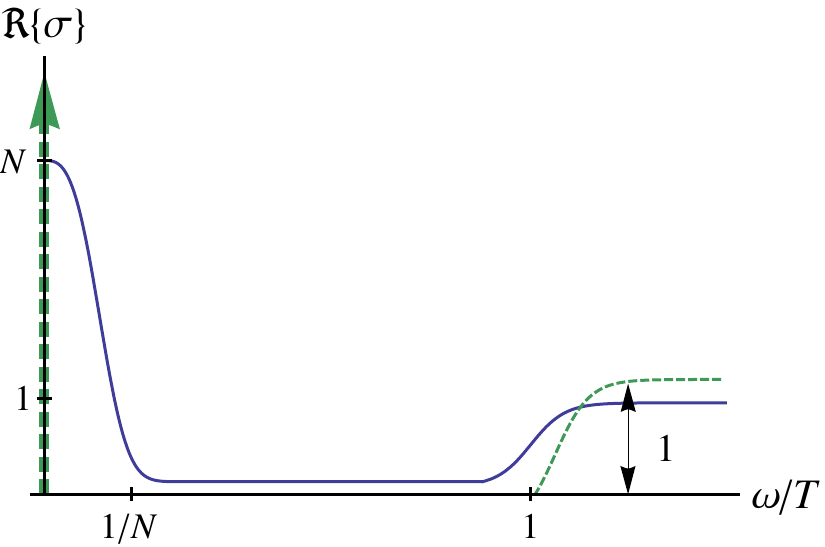}}
 \subfigure[]{\label{fig:sketch-holog}\includegraphics[scale=.6]{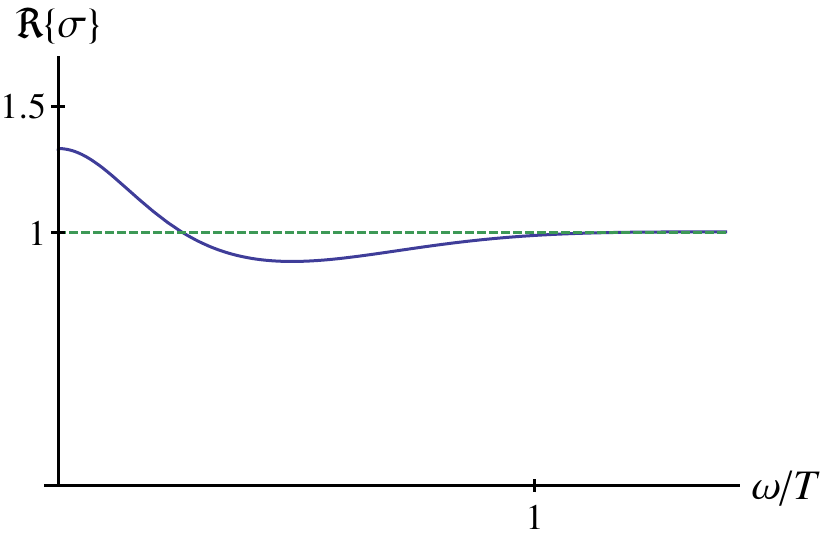}}
\caption{\label{fig:big-pic} 
(a) Perspective on approaches to the charge transport properties of strongly interacting CFTs in 2+1 dimension.
The quantum Boltzmann approach applies to the $1/N$ expansion of the $O(N)$ model: its starting point assumes the existence
of weakly interacting quasiparticles, whose collisions control the transport properties.
In the present paper we start from the ``nearly perfect'' quantum liquid obtained in the $N_c \rightarrow \infty$ limit
of a SU($N_c$) super Yang-Mills theory, which has no quasiparticle description. Holographic methods then allow
expansion away from this liquid ($\lambda$ is the 't Hooft coupling of the gauge theory).
(b) Structure of the charge conductivity in the quantum Boltzmann approach. The dashed line is the $N=\infty$ result:
it has a delta function at zero frequency, and a gap below a threshold frequency. The full line shows the changes from $1/N$ corrections. 
(c) Structure of the charge conductivity in the holographic approach. The $N_c = \infty$ result is the dashed line, and this
is frequency {\em independent}. The full line is the conductivity obtained by including four-derivative terms in the effective
holographic theory for $\gamma > 0$.
} 
\end{figure}

A typical example of such Boltzmann studies is the theory of transport at the quantum critical point of the $N$-component $\phi^4$ field theory with $O(N)$ symmetry
in 2+1 dimensions; the $N=2$ case describes the superfluid-insulator transition of Ref.~\onlinecite{endres}. 
Conformal symmetry emerges at the quantum critical point and the corresponding conformal field theory (CFT) admits a finite
d.c.\ charge conductivity even in the absence of translation-symmetry breaking perturbations\cite{damle} (such as disorder or Umklapp
scattering). This property follows from the presence of independent positive and negative charge excitations related
by charge conjugation (particle-hole) symmetry, which does not require conformal invariance. We shall however
restrict oursevles to CFTs in the current work. The Boltzmann analysis of transport was applied  
in the large $N$ limit of the $O(N)$ model,\cite{ssqhe,sachdev-book,will} 
and the structure of the frequency-dependence of the conductivity, $\sigma (\omega)$, is illustrated in \rfig{sketch-rotor}.
The low frequency behavior is as expected for weakly interacting quasiparticles: there is a Drude peak whose height diverges as $\sim N$,
and whose width vanishes as $1/N$, while preserving the total weight as $N\rightarrow \infty$. It is not at all clear whether such a description of
the low frequency transport is appropriate for the $N=2$ of experimental interest: while it is true that the anomalous exponent $\eta$ remains
small even at $N=2$, it is definitely not the case that the thermal excitations of the quantum critical point interact weakly with each other.
At high frequencies, $\omega \gg T$ ($T$ is the temperature), the predictions of the large $N$ expansion for $\sigma (\omega)$ seem more
reliable: the result asymptotes to a non-zero universal constant $\sigma_\infty$ whose value can be systematically computed order-by-order in the 
$1/N$ expansion, without using the Boltzmann equation. 

\begin{figure}
  \centering
  \includegraphics[scale=.6]{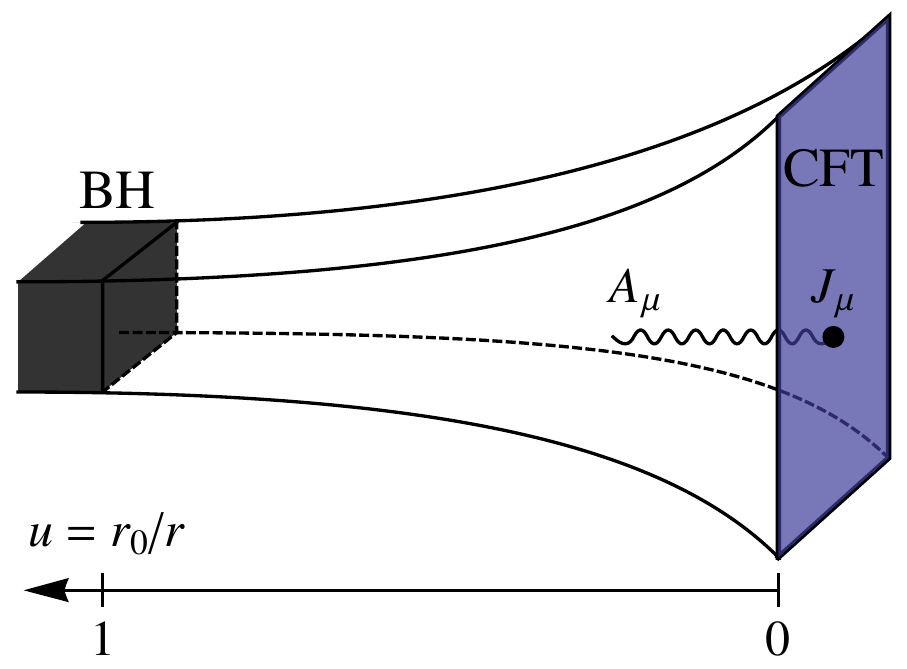}
  \caption{AdS spacetime with a planar black brane. The current ($J_\mu$) correlators of the CFT are related to those
  of the U(1) gauge field ($A_\mu$) in the AdS (bulk) spacetime. The temperature of the horizon of the black brane
   is equal to the temperature of the CFT. The horizon acts as a ``leaky'' boundary to the bulk $A_\mu$ normal modes, which consequently become quasi-normal modes with complex frequencies. These quasi-normal modes specify the finite temperature dynamic properties of the CFT.}
  \label{fig:ads}
\end{figure}

In this paper, we argue for a different physical paradigm as a description of low frequency transport near quantum critical points, replacing the quasiparticle-based
intuition of the Boltzmann equation. We use the description of quantum-critical 
transport based on the AdS/CFT correspondence\cite{m2cft} to emphasize the physical importance of 
``quasi-normal modes'' in the charge response function. Formally, the quasi-normal mode frequencies are the locations of poles in the conductivity in the
lower-half complex frequency plane {\em i.e.\/} the poles obtained by analytically continuing the retarded 
response function from the upper-half plane (UHP)
to the second Riemann sheet in the lower-half plane (LHP). By considering a particle-vortex dual (or ``S-dual'') theory whose conductivity is the inverse of the conductivity of the direct theory, we also associate quasi-normal modes with the poles of the dual theory, which are the zeros of the direct theory. 
Both the pole and zero quasi-normal modes are directly accessible in AdS/CFT methods \cite{star1,star2,hh}, 
and are related to the 
normal modes of excitations in the holographic space: the normal modes have complex frequencies because of the presence of the ``leaky'' horizon of
a black brane; see Fig.~\ref{fig:ads}. 
We will show that a knowledge of these modes allows a complete reconstruction of the frequency dependence of the conductivity, $\sigma (\omega)$,
extending from the hydrodynamic regime with $\omega \ll T$, to the quantum critical regime with $\omega \gg T$. Moreover, these quasi-normal mode
frequencies are also expected to characterize other dynamic properties of the quantum critical system: the recent work of Bhaseen {\em et al.\/} \cite{bhaseen} showed
that the important qualitative features of the approach to thermal equilibrium from an out-of-equilibrium thermal state could be well understood by a knowledge of the structure of the quasi-normal mode frequencies.

Apart from the quasi-normal modes, the long time dynamics also exhibits the well-known \cite{kovtun-rev} classical hydrodynamic feature
of `long time tails' (LTT). The LTT follow from the principles of classical hydrodynamics: arbitrary long wavelength hydrodynamic fluctuations 
lead to the algebraic temporal decay of conserved currents. The LTT depend only upon various transport coefficients, thermodynamic parameters, 
and a high frequency cutoff above which hydrodynamics does not apply. In the quantum-critical systems of interest here, 
this high frequency cutoff is provided by the quasi-normal modes. Thus the LTT describe the dynamics for frequencies $\omega \ll T$,
while the quasi-normal modes appear at $\omega \sim T$ and higher. 
We emphasize that the value of the d.c.\ conductivity, $\sigma(\omega/T=0)$, is determined by the full CFT.
The non-analytic small frequency dependence associated with the LTT can be obtained from the effective 
classical hydrodynamic description which takes the transport coefficients of the CFT treatment as an input.
The focus of the present paper will be on the quasi-normal modes, and we will not have any new results on the LTT; the description of the LTT by holographic methods requires loop corrections to the gravity 
theory \cite{saremi}, which we will not consider here.

From our quasi-normal mode perspective, we will find two exact sum rules that are obeyed by the universal 
quantum critical conductivity, $\sigma (\omega)$,
of all CFTs in 2+1 dimensions with a conserved U(1) charge. These are
\Beq
\int_0^{\infty} d \omega \left[ \Re \, \sigma (\omega) - \sigma_\infty \right] &=& 0,
\label{eq:sum-rule1} \\
\int_0^{\infty} d \omega \left[ \Re \, \frac{1}{\sigma (\omega)} - \frac{1}{\sigma_\infty} \right] &=& 0.
\label{eq:sum-rule2}
\Eeq
Here $\sigma_\infty$ is the limiting value of the conductivity for $\omega \gg T$ (in applications to the lattice models to condensed matter physics,
we assume that $\omega$ always remains smaller than ultraviolet energy scales set by the lattice). 
The first of these sum rules was noted in Ref.~\onlinecite{sum-rules}.
From the point of view of the boundary CFT, Eq.~(\ref{eq:sum-rule1}) is quite natural in
a Boltzmann approach: it is similar to the standard $f$-sum rule,
which we extend to CFTs in Appendix~\ref{app:sum}. There we connect it to an equal-time current correlator,
which we argue does not depend on IR perturbations such as the temperature or chemical potential.
The second sum rule follows from the existence of a 
$S$-dual (or ``particle-vortex'' dual) theory \cite{m2cft,hh,myers11,witten,dopecft} whose conductivity is 
the inverse of the conductivity of the direct
theory. Although it can be justified using the direct sum rule, \req{sum-rule1}, applied to the S-dual CFT,
whose holographic description in general differs from the original theory,
we emphasize that it imposes a further constraint on the original conductivity.
To our knowledge, the second sum rule has not been discussed previously.
All our holographic results here satisfy these two sum rules. % very accurately.
We show in Appendix~\ref{app:ana} that 
the $N=\infty$ result of the $O(N)$ model in Ref.~\onlinecite{damle} obeys the sum rule in \req{sum-rule1}, a feature that was not 
noticed previously. However such quasiparticle-Boltzmann computations do not obey the sum rule
in \req{sum-rule2}.
The holographic computations of the conductivity are the {\em first\/} results which obey not only the sum rule in \req{sum-rule1},
but also the dual sum rule in \req{sum-rule2}.

In principle, the quasi-normal mode frequencies can also be determined by the traditional methods of condensed matter physics.
However, they are difficult to access by perturbative methods, or by numerical methods such as dynamical mean-field theory \cite{dmft}.
One quasi-normal mode is, however, very familiar: the Drude peak of quasiparticle Boltzmann transport, appearing from the behavior $\sigma (\omega) \sim 
\s_0/(1 - i \omega \tau)$, corresponds to a quasi-normal mode at $\omega = - i/\tau$. In a strongly-interacting quantum critical system, we can expect from the arguments of 
Ref.~\onlinecite{damle} that this peak would translate to a quasi-normal mode at $\omega \sim -i T$. As we will see in detail below, 
this single Drude-like quasi-normal mode does not, by itself, provide a satisfactory description of transport, and we need to understand the 
structure of the complete spectrum of quasi-normal modes.
And the most convenient method for determining this complete spectrum is the AdS/CFT correspondence.

As we indicate schematically in \rfig{map}, the AdS/CFT description becomes exact for certain supersymmetric gauge theories in the limit
of a large number of colors, $N_c$, in the gauge group \cite{Maldacena,GKP,Wittenads}. This theory has no quasiparticles, and
in the strict $N_c=\infty$ limit the conductivity is frequency independent even at $T>0$,
as indicated in \rfig{sketch-holog}. Our quasi-normal mode theory expands away from this frequency-independent limit, in contrast to the free particle
limit of the Boltzmann theory (in the latter limit, the Drude contribution becomes $\sigma (\omega) \sim T \delta (\omega)$). 
We describe the basic features of $\sigma(\omega)$ obtained in this manner in the following subsection. Because strong interactions are crucial to the structure
of $\sigma (\omega)$ at all stages, and there is no assumption about the existence of quasiparticles, we expect our results 
to be general description of a wide class of strongly interacting quantum critical points.

\subsection{Generic features of the finite-$T$ conductivity of a CFT}

The frequency dependent conductivity of a CFT in 2+1 dimensions at finite temperature will naturally be a function of the ratio 
of the frequency to the temperature, $\omega/T$,
which we will denote as $w$, with a factor of $4\pi$ convenient in the holographic discussion,
\begin{align}
  w \equiv \frac{\omega}{4\pi T}. \label{defw}
\end{align}
In general, we do not expect the conductivity of a \emph{generic} CFT to be a meromorphic function of
the complex frequency $w$, {\em i.e.\/}  
analytic except possibly at a discrete set of points where it has finite-order poles, all in the LHP. (The latter condition
follows from the causal nature of the retarded current-current correlation function.) The absence of meromorphicity
for the conductivity of an interacting CFT,
or the presence of branch cuts, can be attributed to the LTT \cite{kovtun-rev,kovtun-ltt}.
In the present paper we will not discuss LTT and focus on the meromorphic structure of the conductivity. 
On the one hand, such a description should be valid for CFTs that have a holographic classical gravity description\cite{sum-rules}.
For example, there is strong evidence that certain super Yang-Mills large-$N_c$ gauge theories are holographically dual 
to classical (super)gravity and do not have LTT, which are suppressed by $1/N_c^2$ compared to the leading meromorphic dependence \cite{kovtun-ltt}. 
On the other hand, we believe that understanding the meromorphic structure is a first step to understanding the full analytic structure
of generic CFTs, and do not expect branch cuts from the LTT to significantly modify the poles and zeros of the quasi-normal modes
at frequencies of order $T$ or larger.

\begin{figure}[h]
  \centering
\subfigure[]{\label{fig:analytic-struct-hol}\includegraphics[scale=.51]{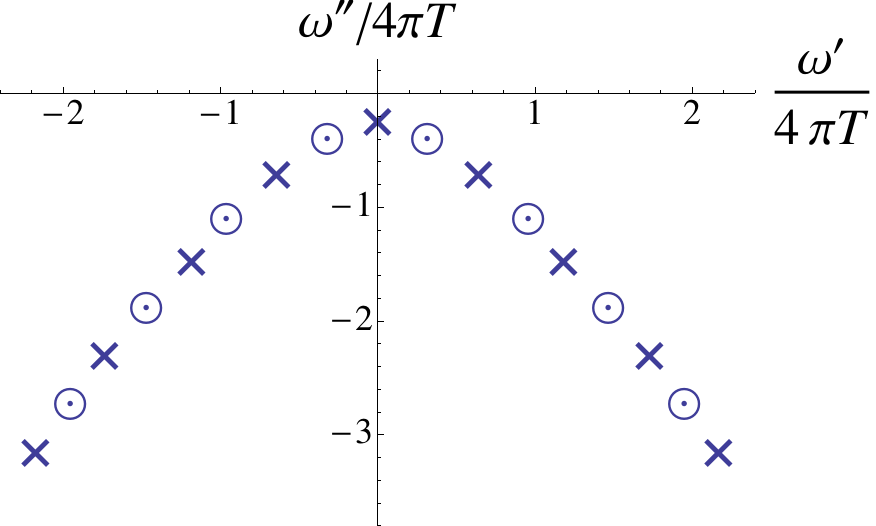}} \quad
\subfigure[]{\label{fig:sig-hol} \includegraphics[scale=.62]{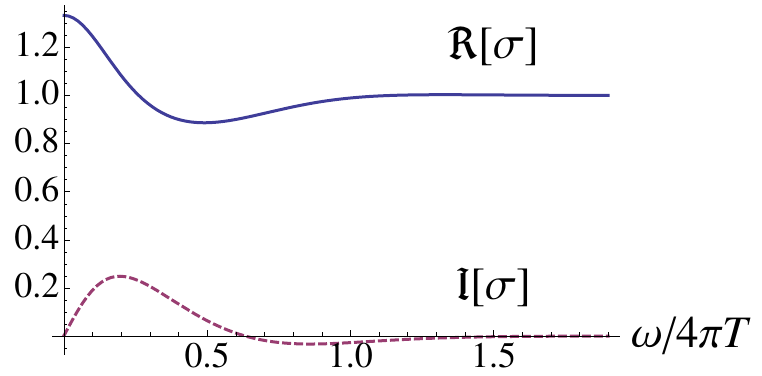}}\\
\subfigure[]{\label{fig:analytic-struct-O(N)} \includegraphics[scale=.6]{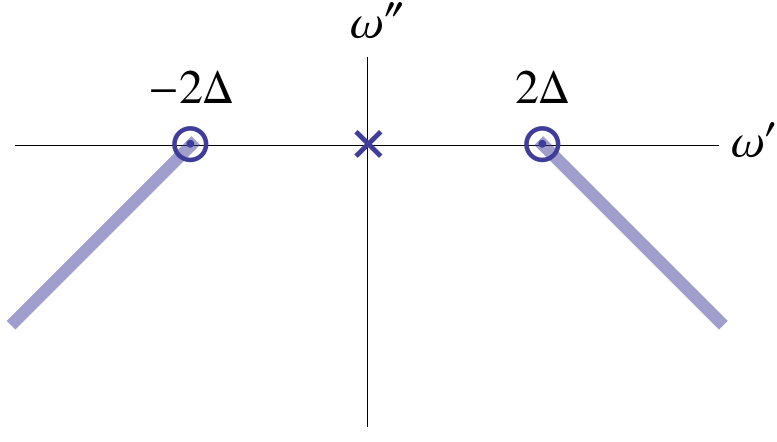}} \quad
 \subfigure[]{\label{fig:sig-full-O(N)}\includegraphics[scale=.65]{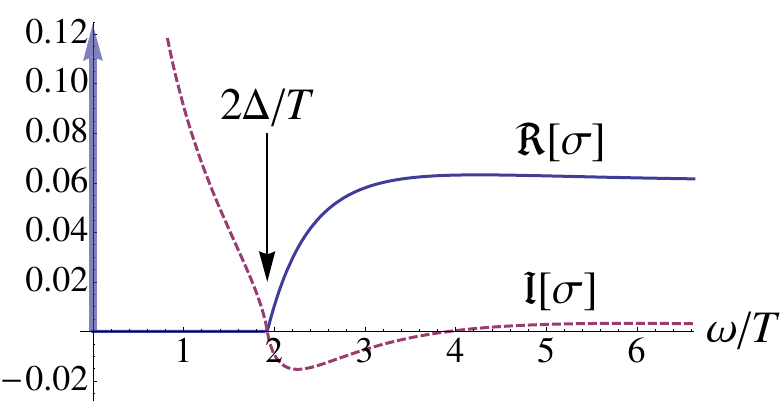}}   
 \caption{\label{fig:summary}
 (a) Poles (crosses) and zeros (circles) of the holographic conductivity at $\gamma = 1/12$. (b) Real and imaginary parts of the holographic conductivity on 
 the real frequency axis. (c) Poles and zeros of the $O(N)$ model at $N=\infty$; the zeros coincide with branch points, and the associated branch cuts
 have been chosen suggestively, indicating that the branch cuts transform into lines of poles and zeros after collisions have been included.
 (d) Conductivity of the $O(N)$ model at $N=\infty$; note the delta function in the real part at $\omega = 0$, and the co-incident zero in
 both the real and imaginary parts at $\omega = 2 \Delta$. In these figures $\Delta/T = 2 \ln ((\sqrt{5}+1)/2)$, and the $O(N)$ computation is reviewed
 in Appendix~\ref{app:ana}.
 }
\end{figure}

The meromorphic condition is tantamount to assuming that in response to a small perturbation, the system will relax exponentially fast to equilibrium
at finite temperature. In addition to LTT, we expect deviations from such behavior to occur at a thermal phase transition for instance,
where power law relaxation will occur. In that case $\s$ is not expected to be meromorphic and branch cuts can appear. 
Another exception is free CFTs, such as the $O(N)$ model in the limit where $N\ra\infty$, where we find poles and zeros
directly on the real frequency axis, as well as branch cuts, as shown in \rfig{analytic-struct-O(N)}.
We restrict ourselves to the finite temperature regime of an interacting conformal quantum critical point with 
a classical gravity description and do not
foresee deviations from meromorphicity\cite{sum-rules}. 

Moreover, we expect the universal conductivity to go to a constant as $w\rightarrow \infty$ \cite{damle,girvin}: 
\begin{align}
  \s(w\rightarrow \infty)=\s_\infty<\infty \,, \quad w\in \mathbb R\,.
\end{align}
%(Note that we are talking about the universal part.) 
Such a well-defined limit will generally not exist as one approaches complex infinity along
certain directions in the LHP.
This is tied to the fact that
$\s$ will not necessarily satisfy the stronger condition of being additionally meromorphic at infinity.
In other words, $s(z):=\s(1/z)$ is not necessarily meromorphic
in the vicinity of the origin, $z=0$. If it were, $\s(w)$ would be a rational function, the ratio
of two finite-order polynomials, and would have a finite number of poles (and zeros).
In our analysis, we shall encounter a class of CFTs whose conductivity
has an infinite set of simple poles, and is thus not meromorphic on the Riemann sphere $\mathbb C\cup \{\infty\}$.
A familiar example of such a function is the Bose-Einstein distribution, $n_B(w)=1/(e^w-1)$, which is meromorphic,
but not at infinity because it has a countably infinite set of poles on the imaginary axis. 
In fact $n_B(1/z)$ has an essential singularity at $z=0$.

A further generic property that $\s$ satisfies in time-reversal invariant systems
is reflection symmetry about the imaginary-frequency axis: $\s(-w^*)=\s(w)^*$, which reduces to evenness or oddness for 
the real and imaginary parts of the conductivity at real frequencies, respectively.
In particular, this means
that all the poles and zeros of $\s$ either come in pairs or else lie on the imaginary axis.
Following this discussion, we can express the conductivity as
\begin{align}
  \s(w)= \frac{\prod{\rm zeros}}{\prod{\rm poles}}=
\frac{\prod_l (w-\zeta_l^0)}{\prod_p (w-\pi_p^0)}\times \frac{\prod_n (w-\zeta_n)(w+\zeta_n^*)}{\prod_m (w-\pi_m)(w+\pi_m^*)}\,,
\end{align}
where $\zeta$ denotes zeros and $\pi$ poles; $\{\zeta_l^0,\pi_p^0 \}$ and  $\{\zeta_m,\pi_m\}$ lie on and off the imaginary axis, respectively. 
In this sense the poles and zeros contain the essential data of the conductivity. 
Actually, since $\s(w\rightarrow\infty)/\s_\infty=1$ on the real axis, which also holds
for all directions in the UHP, they entirely determine $\s/\s_\infty$. In the current holographic analysis, all the poles
and zeros are simple, excluding double (except at special points in parameter space where
these occur) and higher order poles. We suspect this is a general feature of correlated CFTs.
If one is interested in the behavior on the real frequency axis only, the expression for the conductivity
arising from the AdS/CFT correspondence can be truncated to a finite number of poles and zeros: we will show in Section~\ref{sec:truncate}
that this leads to reasonable approximations to the conductivity on the real frequency axis. Such a truncated form
can be compared with experimentally or numerically measured conductivities for systems described by a conformal quantum 
critical point.

As we will show in this paper, the holographic methods allow easy determination of the poles in the conductivity,
which are identified as the frequencies of the quasi-normal modes of the theory on AdS$_4$ in the presence
of a horizon at a temperature $T$. Moreover, the zeros in the conductivity emerge as the frequencies of the
quasi-normal modes of a $S$-dual (or ``particle-vortex'' dual) theory \cite{m2cft,hh,myers11,witten,dopecft}.
We summarize our holographic results for a particular parameter value in Fig.~\ref{fig:summary},
along with the corresponding results for the $O(N)$ model at $N=\infty$.
The $O(N)$ model has a pole at $\omega = 0$, corresponding to the absence of collisions in this model at $N=\infty$. 
This turns into
a Drude-like pole on the imaginary axis, closest to the real axis in the holographic result. We show in Appendix~\ref{app:ana}
that the $O(N)$ model also has a pair of zeros on the real axis, and this is seen to correspond to zeros just below the real
axis in the holographic result. Finally, the $O(N)$ model has a pair of branch points on the real axis; the 
location of the branch cuts emerging from these branch points depends on the path of analytic continuation from the upper half plane.
We have chosen these branch cuts in a suggestive manner in \rfig{analytic-struct-O(N)}, so that they correspond to the lines
of poles and zeros in the lower-half plane of the holographic result. So we see a natural and satisfactory evolution from the analytic
structure of the collisionless quasiparticles of the $O(N)$ model, to the quasi-normal modes of the strongly interacting holographic model.

The outline of our paper is as follows. The holographic theory on AdS$_4$ will be presented in Section~\ref{sec:ads4}.
We will use the effective field theory for charge transport introduced in Ref.~\onlinecite{myers11}, expanded to include
terms with up to 4 spacetime derivatives. The quasi-normal modes will be computed using methods in the literature \cite{star2,hh,hongmcgreevy,denef}.
Section~\ref{sec:qbe} will turn to the traditional quantum Boltzmann methods where new results regarding the analytic structure
are given; in particular, we find that the low frequency Boltzmann conductivity can be accurately represented by a single Drude pole.

\section{Holographic analysis}%: conductivity poles and zeros as quasi-normal frequencies}
\label{sec:ads4}

The AdS/CFT holographic correspondence we use arose from the study of non-abelian supersymmetric gauge theories in the limit of 
a large number of colors, for example with gauge group SU$(N_c)$, $N_c\ra\infty$. By taking an appropriate limit for 
the gauge coupling, such theories are strongly interacting yet they can be described by weakly coupled gravity in an Anti-de-Sitter (AdS)
spacetime with one extended additional spatial dimension, and six or seven compactified ones. 
The fixed-point CFT describing
the strongly correlated gauge theory can be seen as existing on the boundary of AdS. Different correlation functions on the
boundary quantum CFT, such as the charge-current ones of interest to this work, can be computed by using the bulk (semi-)classical 
gravitational theory. For instance, the current operator corresponding to a global U(1) charge in the CFT can be identified with a U(1)
gauge field in the higher dimensional gravitational bulk (\rfig{ads}). We refer the reader to a number of 
reviews\cite{sachdev-book,mcgreevy-rev,sachdev-rev} with condensed matter applications
in mind and proceed to the holographic description of transport in 2+1 dimensional CFTs. 

These CFTs are effectively described by a gravitational bulk theory in 3+1 dimensions. In the case of the supersymmetric
ABJM model\cite{abjm} in a certain limit with an infinite number of colors, the holographic dual is 
simply Einstein's general relativity in the presence of a negative cosmological constant resulting in an AdS$_4$ spacetime.
Charge-transport correlations functions in the CFT can be obtained from those a U(1) probe gauge field with
Maxwellian action in the AdS background. It was shown\cite{m2cft} that the conductivity of the large-$N_c$ ABJM model 
is frequency independent due to an emergent S-duality. Ref.~\onlinecite{myers11} discovered that deviations from self-duality
are obtained by considering 4-derivative corrections to the Einstein-Maxwell theory, which can potentially arise
at order $1/\la$ in the inverse 't Hooft coupling.
The effective action for the bulk gravitational theory discussed in Ref.~\onlinecite{myers11} reads
\begin{align}\label{eq:S_bulk}
  S_{\rm bulk}=\int d^4x \sqrt{-g}\left[\frac{1}{2\ka^2}\left(R+\frac{6}{L^2}\right)
  -\frac{1}{4g_4^2}F_{ab}F^{ab}+\g \frac{L^2}{g_4^2}C_{abcd}F^{ab}F^{cd}\right]\,,
\end{align}
where $g$ is the determinant of the metric $g_{ab}$ with Ricci scalar $R$; $F^{ab}$ is the field strength tensor
of the probe U(1) gauge field $A_a$ holographically dual to the current operator of a global charge of the CFT.
(We use roman indices for the 3+1 spacetime, and greek ones for the boundary 2+1 spacetime.)
Such an action was also considered in Ref.~\onlinecite{ritz}. 
The 4-derivative contribution to charge-transport can be encoded in the last term, proportional to $\g$.
$C_{abcd}$ is the (conformal) Weyl curvature tensor; it is the traceless part of the full Riemann curvature
tensor, $R_{abcd}$: $C_{abcd}=R_{abcd}-(g_{a\left[ c\right.}R_{\left. d\right] b}-g_{b\left[ c\right.}
R_{\left. d\right] a})+\frac{1}{3}R g_{a\left[ c\right.}g_{\left. d\right]b}$.
We observe that the $\g$-term directly couples the probe U(1) gauge field to the metric. 
$L$ is the radius of curvature of the AdS$_4$ space while the gravitational constant $\kappa^2$ is related to 
the coefficient of the two-point correlator of the stress-energy tensor $T_{\mu\nu}$ of the 
boundary CFT (for a review, see Ref.~\onlinecite{suvrat}), an analog of the central charge of CFTs in 1+1D.
The gauge coupling constant $g_4^2 = 1/\sigma_\infty$ dictates the infinite-$w$ conductivity, which we shall set
to 1 throughout, effectively dealing with $\s/\s_\infty$. 
The crucial coupling in this theory is the dimensionless
parameter controlling the four-derivative term, $\gamma$; it determines the structure of a three-point correlator
between the stress-energy tensor and the conserved current.
Stability constraints in the theory imply\cite{myers11} that $|\gamma | \leq 1/12$, and we explore the full range of allowed $\gamma$ values here. Positive values of $\gamma$ yield a low-frequency peak in the conductivity as shown in \rfig{sketch-holog} or \rfig{sig-exp}, 
while negative values of $\gamma$ give rise to a low-frequency dip illustrated in \rfig{sig-dual-exp}, as may be expected from a theory of weakly interacting vortices. Explicit computations of $\gamma$ directly from the CFT yield values \cite{suvrat} in line with these expectations.

In the spirit of the effective field theory approach of Ref.~\onlinecite{myers11}, we should also consider adding other
terms to \req{S_bulk} involving fields other than $F_{ab}$ and the metric tensor \cite{JMthanks}. 
The most important of these are possible ``mass'' terms which tune
the CFT away from the critical point at $T=0$. Such terms are not present in the CFT at $T=0$,
but their values at non-zero $T$ are precisely such that the expectation value of the mass
operator does not change: {\em e.g.\/} in the quantum critical $O(N)$ model of Appendix~\ref{app:sum}, 
$\langle \hat{\phi}_{\alpha}^2 \rangle$ is $T$-independent \cite{CSY}. 
The mass terms can be included in the holographic theory by allowing for 
a scalar dilaton field, $\Phi$, and this can modify charge transport via a term $\sim \Phi F_{\mu\nu} F^{\mu\nu}$.
In the holographic theory, in the absence of external sources, such a dilaton does not acquire an expectation value at $T>0$
when it is not present at $T=0$. And external sources coupling to the gauge field only modify $\Phi$ at quadratic order,
and so $\Phi$ can be neglected in the tree-level linear response. 
Thus even after allowing for additional fields, 
$\gamma$ remains the only important coupling determining the structure of the charge transport at non-zero temperatures.

In the absence of the gauge field, which is here only a probe field used to calculate the linear response, 
the metric that solves the equation of motion associated
with $S_{\rm bulk}$ is:
\begin{align}\label{eq:metric-r}
  ds^2=\frac{r^2}{L^2}\left(-f(r)dt^2+dx^2+dy^2\right) + \frac{L^2dr^2}{r^2f(r)}\,,
\end{align}
where $f(r)=1-r_0^3/r^3$, and $r$ is the coordinate associated with the extra dimension. The CFT
exists on the boundary of AdS, $r\ra\infty$, on the Minkowski spacetime parameterized by $(t,x,y)$.
We emphasize here that the holographic theory is naturally written in real time allowing direct extraction
of the retarded current-current correlation function characterizing the conductivity.
\req{metric-r} corresponds to a 3+1D spacetime with a planar black hole (BH) whose event horizon 
is located at $r=r_0$, and that asymptotically tends to AdS$_4$ as $r\rightarrow\infty$. 
We thus refer to it as Schwarzchild-AdS, or S-AdS. 
The position of the event horizon is directly proportional to the temperature of the boundary CFT,
\begin{align}
  T=\frac{3r_0}{4\pi L^2}\,.
\end{align}
As $T\rightarrow 0$, the black hole disappears and we are left with a pure AdS spacetime, which is
holographically dual to the vacuum of the CFT. The statement that the thermal states of the CFT can
be accessed by considering a BH in AdS can be heuristically understood from the fact
that the BH will Hawking radiate energy that will propagate to the boundary and heat it up.

It will be more convenient to use the dimensionless coordinate $u=r_0/r$, such that \req{metric-r} becomes
\begin{align}
  ds^2=\frac{r_0^2}{L^2u^2}\left(-f(u)dt^2+dx^2+dy^2\right) + \frac{L^2du^2}{u^2f(u)}\,, \qquad f(u)=1-u^3\,.
\end{align}
The boundary, $r=\infty$, is now at $u=0$, while the BH horizon is at $u=1$.

The equation of motion (EoM) for the probe gauge field is the modified Maxwell equation
\begin{align}
  \nabla_a(F^{ab}-4\g L^2 C^{abcd}F_{cd})=0\,,
\end{align}
where $\nabla_a$ denotes a covariant derivative with respect to the background metric, $g_{ab}$. As we are interested in the
current correlator in frequency-momentum space, we Fourier transform the gauge field:
\begin{align}
  A_a(t,x,y,u)=\int \frac{d^3k}{(2\pi)^3} e^{-i\w t+i\b k\cdot\b x}A_a(\w,k_x,k_y,u)\,,
\end{align}
where the coordinate $u$ was left un-transformed since there is no translational invariance
in that direction. We shall actually solve for the full $u$-dependence of $A_a$. We work in the
radial gauge $A_u=0$. Without loss of generality, we also set the spatial momentum to be along the
$x$-direction, $(k_x,k_y)=(k,0)$. In the limit where $k\rightarrow 0$, appropriate 
to a uniform ``electric'' field coupling to the global charge, the equation of motion
for the transverse component, $A_y$, reads 
\begin{align}\label{eq:Ay-ode}
  A_y''+\frac{h'}{h}A_y'+\frac{9w^2}{f^2}A_y=0\,,
\end{align}
where we have defined the dimensionless frequency $w$ in Eq.~(\ref{defw}),
and primes denote derivatives with respect to $u$. The function $h(u)$ is simply $fg$, where $g=1+4\g u^3$
takes the same form as $f=1-u^3$.
As $g(u)$ fully encodes the $\g$-dependence, we wish to make its role more transparent by rewriting the above equation:
\begin{align}\label{eq:AyEoM}
  A_y''+\left(\frac{f'}{f}+\frac{g'}{g} \right) A_y'+\frac{9w^2}{f^2}A_y=0\,.
\end{align}
The term $g'/g=12\g u^2/(1+4\g u^3)$ is seen to be proportional to $\g$, and as such, goes
to zero as $u\rightarrow 0$ consistent with the fact that the Weyl tensor vanishes
in the pure AdS spacetime, which is said to be conformally-flat.

The AdS/CFT correspondence provides an expression for the conductivity of the CFT in terms of the transverse gauge field
auto-correlator evaluated at the boundary, $u=0$,
\begin{align}\label{eq:sig-ads-cft}
  \s(\w)=\left. \frac{i\mc G_{yy}}{\omega} \right|_{u=0}\,,
\end{align}
where $\s(\w)$ is the complex valued conductivity, and $\mc G_{yy}(\w,u)$ is the retarded $A_y$
auto-correlation function. More specifically one gets\cite{m2cft,myers11}:
\begin{align}\label{eq:sig-A}
  \s(w) = \left. -\frac{i}{3w}\frac{\pd_u A_y}{A_y}\right|_{u=0}\,,
\end{align}
where $A_y$ solves the equation of motion \req{AyEoM} with suitable boundary conditions,
as discussed below. The above equation, central to our analysis, has the following heuristic
explanation: $A_y(0)$ acts as a source for the current, while $\pd_uA_y(0)$ is the 
corresponding response. We will see in Section~\ref{sec:qnm} that the quasi-normal modes, {\em i.e.\/} the poles of
conductivity in the LHP, correspond to driving frequencies at which a ``response'' exists in the limit of
vanishing source strength. 

\subsection{Direct solution of conductivity}
The real part of the conductivity on the real frequency axis (retarded correlator) was numerically 
obtained in Ref.~\onlinecite{myers11}. We extend
their analysis from real to complex frequencies, $w\in\mathbb C$. The boundary conditions necessary to
solve \req{AyEoM} are imposed at the BH event horizon\cite{myers11} at $u=1$.  
To obtain them we examine the EoM near the horizon, which admits 
the following two solutions:
$A_y\sim (1-u)^{\pm iw}$. These correspond to outgoing and ingoing waves from the point of view of the BH, respectively.
The retarded correlator is obtained by choosing the ingoing condition. To implement this in the numerical solution,
we factor out the singular behavior: $A_y=(1-u)^{-iw}F(u)$, where $F(u)$ is the sought-after function; it is regular at the horizon. 
From \req{sig-A}, we see that we are free to fix one of the two boundary conditions, either for $A_y(1)$ or $A_y'(1)$,
to an arbitrary finite constant without altering the conductivity. We impose $A_y(1)=F(1)=1$.
The appropriate boundary condition for $F'$ can be obtained by examining the differential equation near $u=1$ as
is discussed in Ref.~\onlinecite{myers11} and in Appendix~\ref{ap:solv-sig}. 
\begin{figure}[h]
\centering%
\subfigure[$\Re\{ \s(w;\g=1/12)\}$]{\label{fig:sp}\includegraphics[scale=.45]{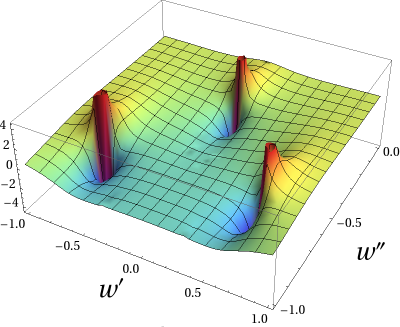}}  
\subfigure[$\Re \{\hat\s (w;\g=1/12)\}$]{\label{fig:shp} \includegraphics[scale=.45]{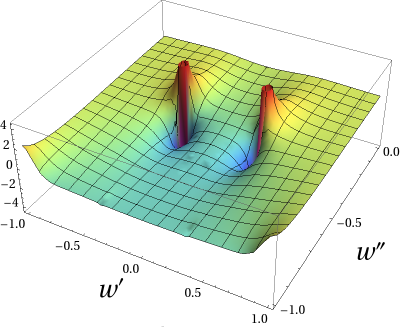}}\\
\subfigure[$\Re\{\s(w;\g=-1/12)\}$]{\label{fig:sm}\includegraphics[scale=.45]{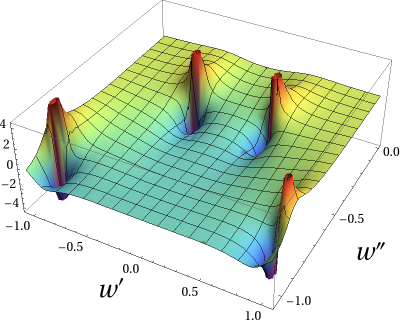}}  
\subfigure[$\Re\{\hat\s(w;\g=-1/12)\}$]{\label{fig:shm} \includegraphics[scale=.45]{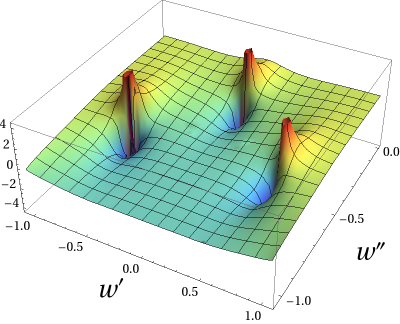}}
\caption{\label{fig:sig-direct} Conductivity $\s$ and its S-dual $\hat\s=1/\s$ in the LHP, $w''=\Im w \leq 0$, for
$|\g|=1/12$. The zeros of $\s(w;\g)$ are the poles of $\hat\s(w;\g)$.
We further note the qualitative correspondence between the poles of $\s(w;\g)$ and the zeros of $\hat\s(w;-\g)$. 
} 
\end{figure}

All the poles of the conductivity are in the LHP, as it is obtained from the 
retarded current-current correlation function. The numerical result is shown in \rfig{sp} and \rfig{sm} for the two
values of $\g$ saturating the stability bound, $\g=\pm 1/12$, respectively. 
\rfig{sp} shows the conductivity for $\g=1/12$, which corresponds to particle-like transport with
a Drude peak at small real frequencies as can be seen on the real $w$-axis, or more clearly in \rfig{sig-hol} or \rfig{sig-exp}. 
Such low-frequency behavior is dictated by a Drude pole, located closest to the origin. The numerical
solution also shows the presence of satellite poles, the two dominant ones being shown.
These are symmetrically distributed about the $\Im w$ axis as required by time-reversal, and are essential to capture
the behavior of $\s$ beyond the small frequency limit. In contrast, the conductivity at $\g=-1/12$ in \rfig{sm} shows a minimum
at $w=0$ on the real axis, see also \rfig{sig-dual-exp} for a plot restricted  to real frequencies. The corresponding
pole structure shows no poles on the imaginary axis, in particular no Drude pole. 
%The poles, such as the 4 in \rfig{sm},
%can be naturally guessed from the oscillations of the conductivity at real frequencies. 
The conductivity at $\g=-1/12$ is said to be vortex-like because it can be put in correspondence with the 
conductivity of the CFT S-dual to the one with $\g=1/12$,
as we now explain.

\subsection{S-duality and conductivity zeros}

Great insight into the behavior of the conductivity can be gained by means of S-duality, a generalization of the
familiar particle-vortex duality of the O$(2)$ model. S-duality on the boundary CFT is mirrored by electric-magnetic (EM) duality
for the bulk U(1) gauge field, which we now briefly review. Given the abelian gauge theory for the U(1) bulk field
$A_a$, we can always perform a change of functional variables in the partition function to a new gauge field $\hat A_a$
by adding the following term to $S_{\rm bulk}$, \req{S_bulk}:
\begin{align}
  S'=\int d^4x\sqrt{-g}\,\frac{1}{2}\ve^{abcd}\hat A_a\pd_bF_{cd}\,,
\end{align}
with the corresponding functional integral for $\hat A_a$. Performing the integral over $\hat A_a$ would simply
enforce the Bianchi identity, $\ve^{abcd}\pd_bF_{cd}=0$, implying $F_{ab}=\pd_aA_b-\pd_bA_a$, where $\ve_{abcd}$ is the
fully-antisymmetric tensor in 3+1D with $\ve_{txyu}=\sqrt{-g}$. 

If instead one integrates out $A_a$ first, a new action in terms of $\hat A_a$ results:
\begin{align}
  \hat S_{\rm bulk}=-\int d^4x\sqrt{-g}\,\frac{1}{8\hat g_4^2}\hat F_{ab}\hat X^{abcd}\hat F_{cd}\,,
\end{align}
where we have defined the field strength of the dual gauge field, $\hat F_{ab}=\pd_a\hat A_b-\pd_b\hat A_a$,
and dual coupling $\hat g_4=1/g_4$.
An exactly analogous action holds for $A_a$ without the hats. The rank-4 tensors $X,\hat X$ are
shorthands to simplify the actions:
\begin{align}
  X_{ab}{}^{cd}&=I_{ab}{}^{cd}-8\g L^2C_{ab}{}^{cd}\,, \\
 \hat X_{ab}{}^{cd}&=\frac{1}{4}\ve_{ab}{}^{ef}(X\inv)_{ef}{}^{gh}\ve_{gh}{}^{cd}\,, 
\end{align}
with the rank-4 tensor $I_{ab}{}^{cd}\equiv\de_a{}^c\de_b{}^d-\de_a{}^d\de_b{}^c$, the identity on the
space of two-forms, for e.g. $F_{ab}=\tfrac{1}{2}I_{ab}{}^{cd}F_{cd}$. The inverse tensor of $X$ is then
defined via $\tfrac{1}{2}(X\inv)_{ab}{}^{cd}X_{cd}{}^{ef}=I_{ab}{}^{ef}$. In terms of the $X$-tensors, 
the EoM for $A_a$ and $\hat A_a$ simply read:
\begin{align}
  \nabla_b(X^{abcd}F_{cd})=0\,,\\
  \nabla_b(\hat X^{abcd}\hat F_{cd})=0\,. \label{eq:X-EoM-dual}
\end{align}
It can be shown\cite{myers11} that for small $\g$, the dual $X$-tensor has the following Taylor expansion:
\begin{align}
  \hat X_{ab}{}^{cd}&=I_{ab}{}^{cd}+8\g L^2C_{ab}{}^{cd}+\mc O(\g^2)\,,\\
  &= X_{ab}{}^{cd}\left.\right|_{\g\ra-\g}+\mc O(\g^2)\,.
\end{align}
We thus see that if $\g=0$, $X=\hat X$ and the actions, and associated EoM, for $A$ and $\hat A$ have the same
form. In that case, the two theories are related by an exchange between electric and magnetic fields:
the standard EM (hodge) \emph{self-duality} of electromagnetism. In contrast, in the presence of the 4-derivative
term parameterized by $\g$, the EM self-duality is lost. However, at small $\g$ the EM duality is particularly
simple and will serve as a guide for any finite $\g$: the holographic theory for $\g$ maps to the one
for $-\g$, neglecting $\mc O(\g^2)$ contributions. 

Let us now examine the impact of this bulk EM duality, $A\ra \hat A$, on the boundary CFT. The holographic
correspondence relates the bulk gauge field $A$ to the current of a global U(1) charge
of the CFT, $J$. In the same way, the dual gauge field $\hat A$ will couple to the 
current $\hat J$ of the S-dual CFT, which generically \emph{differs} from the original CFT. 
Just as the conductivity of the original CFT, $\s$, is related to the $J$ auto-correlator, the conductivity
of the S-dual CFT, $\hat\s$, will be obtained from the $\hat J$ auto-correlator. The conductivities of the
S-dual CFT pair are in fact the inverse of each other:
\begin{align}\label{eq:sig-dual}
  \hat \s(w;\g)=\frac{1}{\s(w;\g)}\,,
\end{align}
where we emphasize that this relation holds for the complex conductivities, $\s=\Re\s+i\Im\s$. 
We present the short proof here using results of Ref.~\onlinecite{myers11}. (We note that
such a result was derived for a specific class of CFTs in Ref.~\onlinecite{m2cft}.)
We begin with the general form of the retarded current-current correlation function: 
$\mc G_{\mu\nu}(\w,\b q)=\sqrt{q_\la q^\la}(P_{\mu\nu}^TK^T(\w,q)+P_{\mu\nu}^TK^L(\w,q))$, with the orthogonal
transverse and longitudinal projectors $P^{T,L}$: $P_{tt}^T=P_{ti}^T=P_{it}^T=0$, $P_{ij}^T=\de_{ij}-q_iq_j/q^2$,
and by orthogonality: $P_{\mu\nu}^L=[\eta_{\mu\nu}-q_\mu q_\nu/(q_\la q^\la)]-P_{\mu\nu}^T$. The Minkowski
metric was introduced, $\eta_{\mu\nu}=\diag(-1,1,1)$, such that $q_\la q^\la=\eta_{\la\la'}q^{\la}q^{\la'}=-\w^2+q^2$.
Of interest to us is the holographic relation between the transverse correlator giving the 
conductivity and the bulk gauge field correlator, $\mc G_{\mu\nu}$: 
\begin{align}\label{eq:rel-K-G}
  \sqrt{q^2-\w^2}K^T(\w,q)=\left.\mc G_{yy}(\w,q)\right|_{u=0}=\w\s(\w,q)/i\,,
\end{align}
where $\s(\w,q)$ is the frequency and momentum dependent conductivity.
The same expression (with hats) holds in the S-dual theory. Using the
action of EM duality on the bulk, Ref.~\onlinecite{myers11} showed the relation:
\begin{align}
  K^T(\w,q)\hat K^L(\w,q)=1\,,
\end{align}
that relates the transverse current-current correlator of the original CFT to the longitudinal one of the dual CFT.
When combined with the fact that in the limit of vanishing spatial momentum, $q\ra 0$, rotational invariance enforces
$K^T(\w,q)=K^L(\w,q)$, which is also naturally true with hats, we obtain
\begin{align}
  \hat K^T(\w,q=0)=\frac{1}{K^T(\w,q=0)}\,.
\end{align}
By virtue of \req{rel-K-G} and its dual version, this concludes the proof of \req{sig-dual}.

The poles of the dual conductivity, $\hat\s=1/\s$, then must correspond to the zeros of the conductivity, $\s$,
and vice versa. 
As a consequence, we see that \emph{S-duality interchanges the locations of the 
conductivity zeros and poles}.
This is is consistent with the direct solution shown in \rfig{sig-direct}.
Take for example the theory at $\g=1/12$, \rfig{sp}: it will have a Drude pole on the imaginary axis, which gives rise 
to a Drude peak at small frequencies. Under S-duality this pole becomes a \emph{Drude zero} of $\hat \s$, \rfig{shp}, and 
the conductivity of the new theory will have a minimum at small frequencies. 

As we saw above, changing the sign of $\g$ corresponds to an approximate S-duality valid for $|\g|\ll 1$.
More generally, in terms of the ``pole/zero-topology'' or ordering, both operations are equivalent. Indeed, if we consider
the pole/zero structure of the positive frequency branch of the conductivity $\Re w\geq 0$ 
(which is sufficient by time-reversal)
and order the poles and zeros according to their norm, we get the following two equivalence classes:
\begin{align}
  \rm {\bf pole}&\rm -zero-pole-zero-\dots\quad \ra \quad particle\text{\,-}like\;  (\g >0\; for\; e.g.)\,,\\
  \rm {\bf zero}&\rm -pole-zero-pole-\dots\quad \ra \quad vortex\text{-}like \;  (\g <0\; for\; e.g.)\,,
\end{align} 
where the first label (in bold) designates the Drude pole or zero. In the above, we also 
designate two consecutive poles/zeros by ``pole/zero''. Such a situation occurs when two poles/zeros become 
bound to the imaginary frequency axis at sufficiently small $\g$, as seen in \rfig{zip}. (The same
caveat is applicable to the special values of $\g$ at which pairs of poles/zeros become bound to the
imaginary axis and form a double pole/zero.) 
Note that the leading pole or zero always comes alone.
Both S-duality and $\g\ra-\g$ interchange
these two analytic structures. 
This underlies the qualitative correspondence between the pole structure of $\s(w;\g)$ and that of
$\hat\s(w;-\g)$; for example, compare \rfig{sp} and \rfig{shm}, or \rfig{sm} and \rfig{shp}.
The correspondence quantitatively improves in the limit of small $\g$. Explicitly, 
\begin{align}\label{eq:2-s-dualities}
  \s(w;\g)\approx\frac{1}{\s(w;-\g)}\,, \quad |\g|\ll 1\,,
\end{align}
holds because performing $\s\ra 1/\s$ together with $\g\ra-\g$ is approximately tantamount to two S-duality transformations and
is equivalent to the identity, modulo $\mc O(\g^2)$ terms. 

Finally, we mention that for a given $\g$ it is not possible to find a $\g'$ such that $\hat\s(w;\g)=\s(w;\g')$.
In other words, the dual of the boundary CFT with parameter $\g$ cannot correspond to the original CFT with a different
parameter $\g'$. This can be seen as follows. We first require that the relation hold true at zero frequency: 
$\hat\s(0;\g)=\s(0;\g')$, which implies
$1/(1+4\g)=1+4\g'$ or $\g'=(\tfrac{1}{1+4\g}-1)/4$, where we have used $\s(0;\g)=1+4\g$ (see Refs.~\onlinecite{myers11,ritz}). 
Although for this value of $\g'$, $\hat\s(w;\g)$ and $\s(w;\g')$
agree for both $w,1/w=0$, we have numerically verified that they always disagree at intermediate frequencies,
the disagreement decreasing as $\g\ra 0$, in which limit $\g'\approx -\g$. The absence of a $\g'$ satisfying $\hat\s(w;\g)=\s(w;\g')$
is in accordance with the fact that holgraphic action of the S-dual CFT contains terms beyond $C_{abcd}F^{ab}F^{cd}$. The latter
is only the first term in the Taylor expansion in $\g$.

We now turn to a better method of determining the poles and zeros, as
the direct solution of \req{Ay-ode} can only reliably capture the poles nearest to the origin. 
The main problem with the direct solution of the differential equation for $A_y$, \req{Ay-ode},
is that the Fourier modes $A_y(u;w)$ at the UV boundary, $u=0$, generically grow exponentially as the
imaginary part of the frequency $\Im w$ becomes more and more negative making the numerical results
unstable. Although an exception occurs at the poles, where $A_y(u=0;\w_{\rm pole})$ vanishes (see below),
it is hard to untangle the true analytical structure from the numerical noise, 
hence the need for a more sophisticated approach.

\subsection{Quasi-normal modes and poles}
\label{sec:qnm}
\begin{figure}
\centering%
\subfigure[$\s(\g=1/12)$]{\label{fig:pure-rot3}\includegraphics[scale=.35]{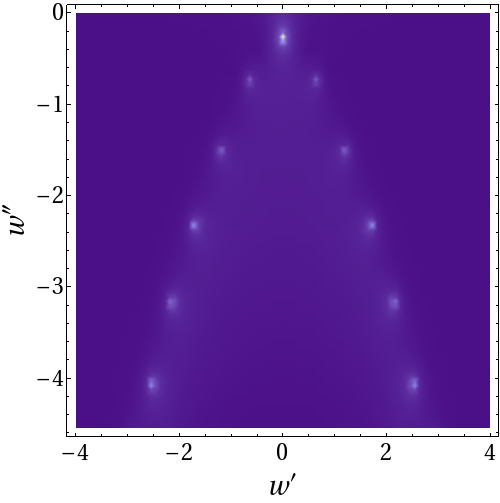}}  
 \subfigure[$\hat\s (\g=1/12)$]{\label{fig:gauged-rot3} \includegraphics[scale=.35]{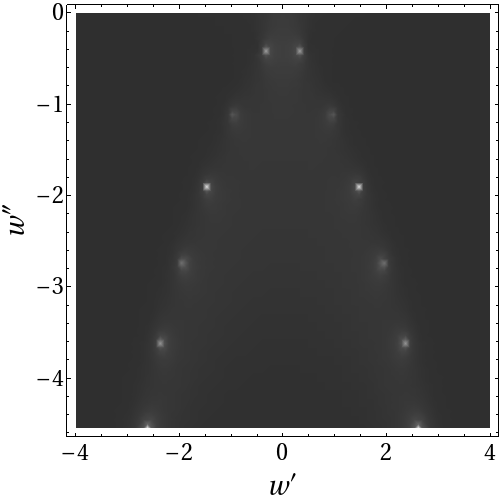}}\\
 \subfigure[$\s(\g=-1/12)$]{\label{fig:gauged-rot4} \includegraphics[scale=.35]{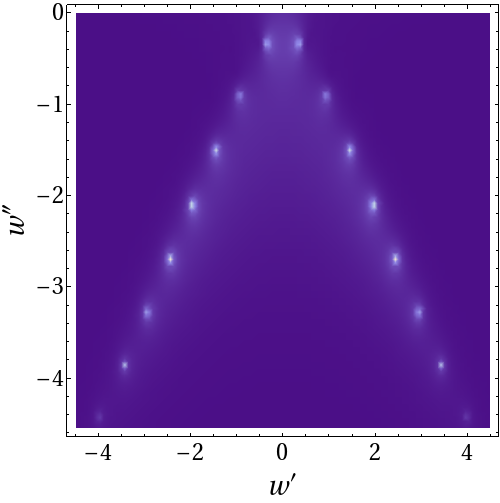}}
 \subfigure[$\hat \s(\g=-1/12)$]{\label{fig:pure-rot4}\includegraphics[scale=.35]{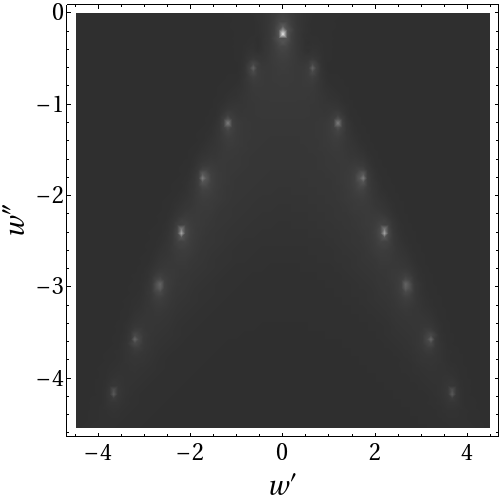}}
\caption{\label{fig:qnm} Quasi-normal modes (bright spots) of the transverse gauge mode for $\g=|1/12|$ 
in the complex frequency plane, $w=w'+iw''$.
The QNMs correspond to the poles of the conductivity (a \& c). EM duality yields the QNMs of the
dual gauge mode, and these correspond to the poles of the dual conductivity, $\hat \s(w)=1/\s(w)$,
i.e. the zeros of $\s(w)$, see panels b \& d.
} 
\end{figure}

We present an alternative and more powerful method of capturing the poles by considering the so-called 
quasi-normal modes (QNMs) of the gauge field in the curved S-AdS$_4$ spacetime. 
These modes are eigenfunctions of the EoM for $A_y$, \req{Ay-ode}:
\begin{align}\label{eq:qnm}
  \mc A_n''+\frac{h'}{h}\mc A_n'+\frac{9w_n^2}{f^2}\mc A_n=0\,,
\end{align}
where $\mc A_n$ is a QNM with frequency $w_n$. The QNM have the special property that they
vanish at the boundary: $\mc A_n\rightarrow 0$ as $u\rightarrow 0$. From the expression for the conductivity,
\req{sig-A}, we can see that this will lead to $w_n$ being a singular point of the conductivity:
\begin{align}
  \s(w_n)\sim \left. \frac{\pd_u \mc A_n}{\mc A_n}\right|_{u=0} \sim \frac{\pd_u \mc A_n(0)}{0}\ra\pm\infty\,,
\end{align}
where $\pd_u \mc A_n(0)$ is generically finite at the QNM frequencies where $\mc A_n(0)=0$.
[In contrast, the conductivity zeros or QNM of the EM-dual Maxwell equation correspond to
frequencies at which $\pd_u \mc A(0)=0$ but $\mc A(0)$ is finite.] 
The name quasi-normal instead of normal is used
because the eigenfunctions $\mc A_n$ diverge approaching the BH horizon, $u=1$. This follows from
the above-mentioned asymptotic form near the horizon, $\mc A_n\sim (1-u)^{-iw_n}=(1-u)^{w_n''-iw_n'}$,
implying a divergence for frequencies in the LHP. 
As predicted by the AdS/CFT correspondence and
verified by our numerical analysis, shown in \rfig{qnm}, the QNMs indeed agree with the
poles of the conductivity shown in \rfig{sig-direct} and more precisely in \rfig{sig2}.
The QNMs are found by using a Frobenius expansion
\begin{align}\label{eq:qnm-exp}
  A_y = u f(u)^{-iw}\sum_{m=0}^M a_m (u-\bar u)^m\,,
\end{align}
where we have factored out the behavior near the event horizon, $f(u)^{-iw}\sim (1-u)^{-iw}$, and 
near the boundary, $u$. We have chosen to Taylor expand around $\bar u=1/2$; $M+1$ is the number of terms in the 
truncated series. Substituting \req{qnm-exp} in \req{Ay-ode} yields a matrix equation for the coefficients, $a_m$: 
\begin{align}
  \sum_{m=0}^M B_{lm}a_m =0\,,
\end{align}
where the l.h.s.\ is the coefficient of $(u-\bar u)^l$, $0\leq l\leq M$. Note that $B_{lm}=B_{lm}(w)$ and 
$a_m=a_m(w)$ both depend on the frequency, and although not explicitly shown, on $\g$ as well. 
For fixed $\g$, this homogeneous system of linear equations
has a solution at a set of frequencies $\{w_n\}$ at which $\det B(w_n)=0$. Or equivalently, when
the smallest-normed eigenvalue of $B$, $\la_{\rm min}$, vanishes, which we find more convenient to implement numerically.
Plots of $1/|\la_{\rm min}|$ (multiplied by an exponential function to improve the visibility) as a 
function of $w$ are given in
\rfig{qnm} for $|\g|=1/12$. The QNMs are the bright spots. 
In obtaining the QNMs of the dual conductivity, $\hat\s=1/\s$, we have used the EoM for the dual
gauge field $\hat A$, \req{X-EoM-dual}:
 \begin{align}\label{eq:Ay-dual}
  \hat A_y''+\left(\frac{f'}{f}-\frac{g'}{g} \right) \hat A_y'+\frac{9w^2}{f^2}\hat A_y=0\,.
\end{align}
It differs from the one for $A_y$, \req{AyEoM}, by the negative sign\cite{myers11}. Note that this shows that 
$\g\ra-\g$ does not exactly correspond to S-duality, because the former would give $-g'/(1-4\g u^3)\neq -g'/g$, where $g=1+4\g u^3$.

Whereas the direct solution only gives reliable
answers up to $\Im w\sim -1$, the QNM approach has a wider range of applicability 
and is numerically more stable giving us more
insight into the analytic structure. We have performed a WKB analysis in Appendix~\ref{app:wkb}
to determine the asymptotic QNMs for $|w|\gg 1$. 
We next examine the transition
that occurs when going from positive to negative values of $\g$.

\subsection{Pole motion and S-duality}
\begin{figure}
\centering
 \subfigure[ $\g=10^{-2}\rightarrow 10^{-3}$]{\label{fig:zip1} \includegraphics[scale=.55]{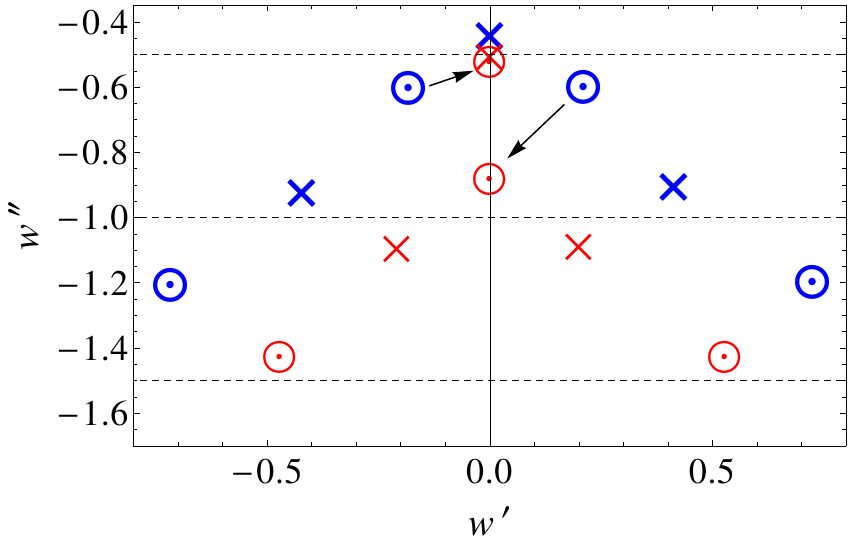}}
 \subfigure[ $\g=10^{-3}\rightarrow 10^{-4}$]{\label{fig:zip2} \includegraphics[scale=.55]{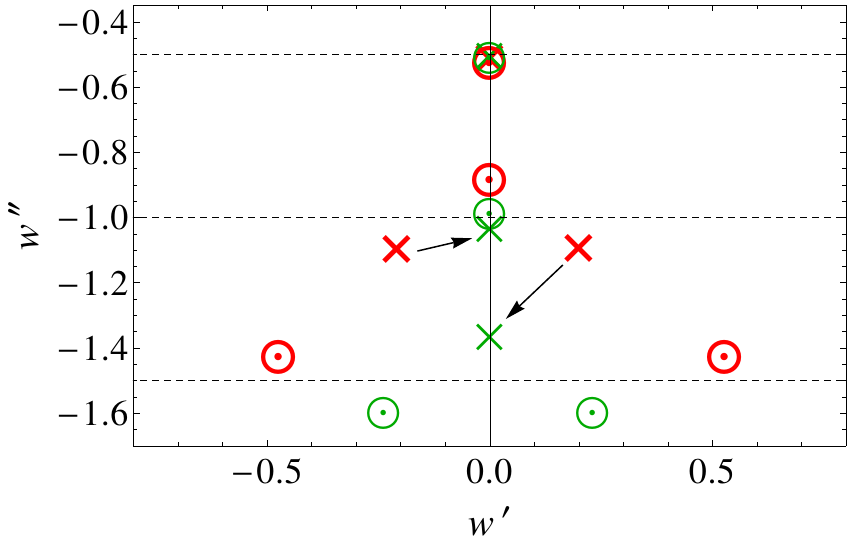}}\\
 \subfigure[ $\g\sim 0^+$]{\label{fig:zipped} \includegraphics[scale=.55]{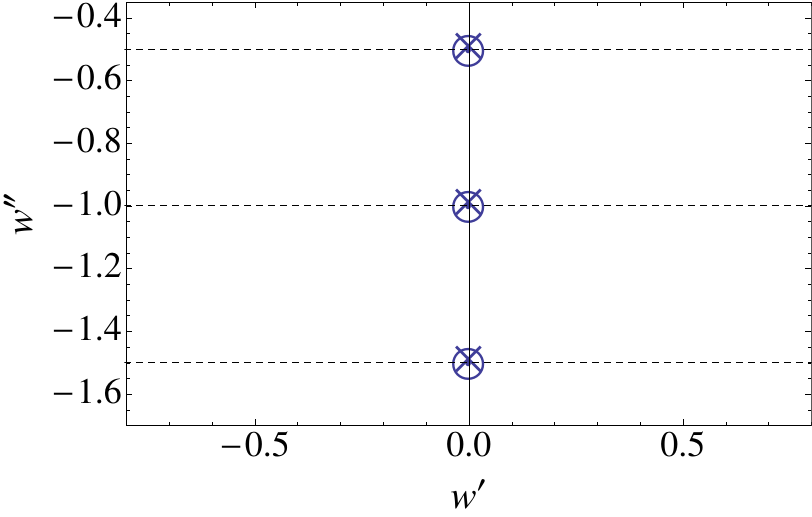}}
 \caption{\label{fig:zip} Illustration of the motion of the poles and zeros as $\g$ goes to zero,
in 3 steps: $\g=10^{-2}\rightarrow 10^{-3}\rightarrow 10^{-4}\ra 0^+$. In each panel the motion is
from bold to thin as $\g$ decreases; with crosses representing 
poles while circles, zeros. a) Blue thick markers are for $\g=10^{-2}$, while the red thin
ones for $\g=10^{-3}$. b) The red and thick markers are for $\g=10^{-3}$, while the green thin
ones for $\g=10^{-4}$. c) ``Zipped'' pole-zero structure for $\g\sim 0^+$, where only poles and zeros far
from the origin will lie off the imaginary axis. 
} 
\end{figure}

The motion of the poles and zeros as $\g$ changes sign is illustrated in \rfig{zip} for $\g>0$. 
For $\g<0$, one simply interchanges the zeros and poles, i.e. the crosses and circles. 
The pole/zero motion can be loosely compared with a ``zipper mechanism''. The arrows in \rfig{zip} show
the non-trivial motion of a pair of poles or zeros as they become ``zipped'' to the imaginary axis.
(A caveat regarding the arrows: by time-reversal symmetry, $w \rightarrow -w^*$, so we cannot say which pole 
goes to which once they become pinned to the imaginary axis. The arrows are just a guide.)
For sufficiently small $\g$, each point on the imaginary axis located at $w_n^{\rm zip} = -i n/2$, 
where $n$ is a positive integer,
will have a pole and zero arbitrarily close to it. When $\g$ = 0, they will ``annihilate'' as it should
because the complex conductivity for $\g=0$ has no poles or zeros as it
takes the constant self-dual value for all complex frequencies. It should be noted that since $w=\w/4\pi T$,
the annihilation frequencies are
\begin{align}
  \w_n^{\rm zip}=-i2\pi n T\,,\;\; n=1,2,3,\dots\,,
\end{align}
i.e. the bosonic Matsubara frequencies in the LHP. Although this results seems natural, we do not
have a clear explanation for it and leave the question for future investigation. 
Finally, from the direct numerical solution of the EoM, we have looked at the residue of the pole 
near $w=-i/2$ (closest to the origin), and have found that it decreases linearly with $\g$,
consistent with the $\g=0$ limit. 

The motion of a pair of poles becoming attached to the imaginary axis bears some similarity to that found
in a recent paper\cite{bhaseen}, 
where as the (dynamic) spontaneous symmetry breaking happens, 
a pair of QNM poles becomes glued to the imaginary axis. In their case one of the poles stays at the origin, 
signaling a gapless Goldstone boson. We will see below one peculiar limit where a conductivity pole hits the origin.

\subsection{Truncations}
\label{sec:truncate}

\begin{figure}[h]
  \centering
\subfigure[]{\label{fig:sig-exp}\includegraphics[scale=.55]{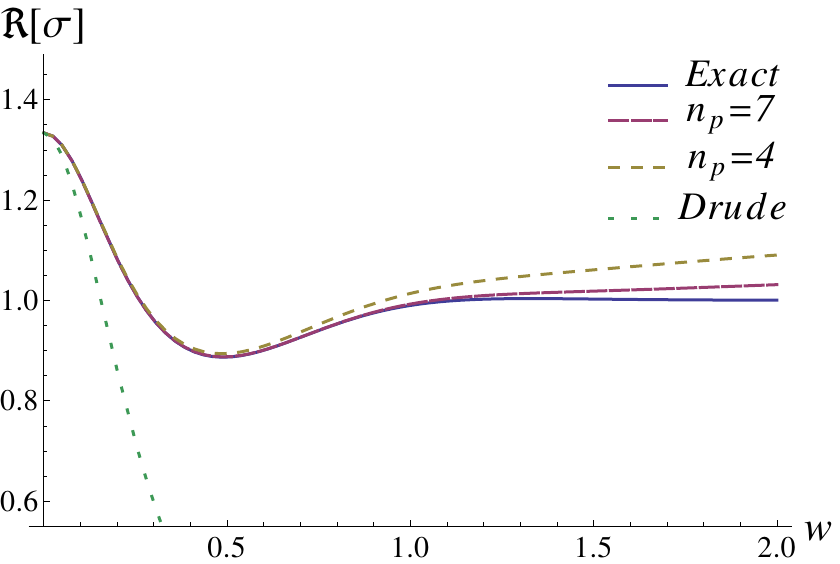}}  
\subfigure[]{\label{fig:sig-dual-exp} \includegraphics[scale=.55]{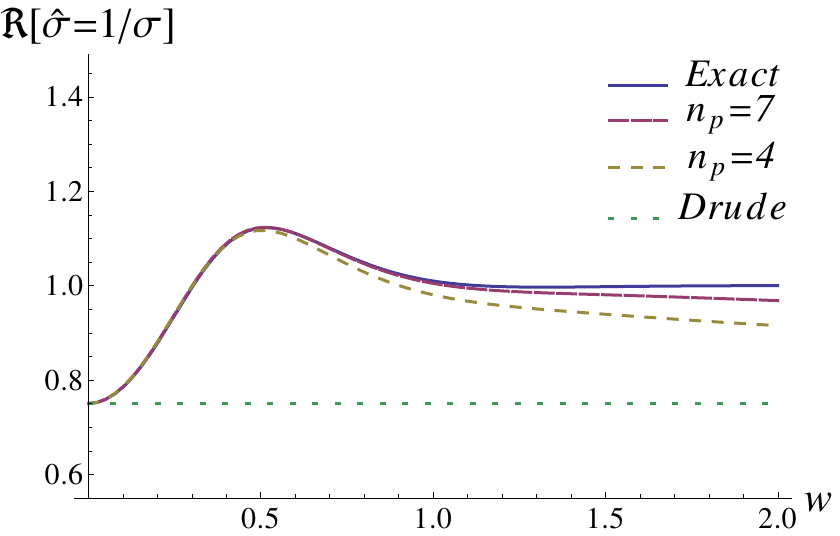}}\\
  \caption{\label{fig:sig-pole-exp}
Conductivity (a), and its dual (b), $\hat\s=1/\s$, arising from a holographic treatment with a truncated number of poles,
$2n_p-1$. One pole lies on the imaginary axis, the Drude pole, while $n_p-1$ pairs have a finite
real part. The Drude form is characterized by a single pole: $\s=\s_0/(1-i\w\tau)$.}
\end{figure}

If one is interested in the behavior on the real frequency axis only, the expression for the conductivity
arising from the AdS/CFT correspondence can be truncated to a finite number of poles and zeros. For instance, in 
a parameter regime believed to be of interest to the a wide class of CFTs, the conductivity has a single
purely imaginary pole, accompanied by satellite poles off the imaginary axis. By truncating the number of poles
we obtain an excellent approximation to the exact dependence as we show in \rfig{sig-exp}: $n_p$ counts the
number of poles/zeros, not counting the time-reversal partners.
The truncated conductivity reads
\begin{align}
  \s_{n_p}(w)=
\frac{(w-\zeta^0)}{ (w-\pi^0)} \prod_{n=1}^{n_p-1}\frac{ (w-\zeta_n)(w+\zeta_n^*)}{(w-\pi_n)(w+\pi_n^*)}
\end{align}
where $2n_p-1$ is the odd number of poles or zeros (the $-1$ follows because the Drude pole/zero is
its own time-reversal partner). The value of the zero $\zeta^0$ is obtained by fixing $\s(0)=\s_0$.
Just like $\pi_0$, it lies on the imaginary axis: 
\begin{align}
  \frac{\zeta_0}{\pi_0}= \s_0\prod_{n=1}^{n_p-1}\left|\frac{\pi_n}{\zeta_n}\right|^2\,.
\end{align}
It is included so that the truncated conductivity goes to a finite constant as $\lim_{w\rightarrow \infty}\s=\s_\infty>0$. 
\rfig{sig-dual-exp} shows the corresponding dual conductivity, $\hat\s(w)=1/\s(w)$, whose poles/zeros
correspond to the zeros/poles of $\s$. Note that the real part of the dual Drude conductivity, $\hat\s=1/\s=(1-w/\pi^0)/\s_0$,
is trivially constant (for real frequencies).

\section{Emergence of Drude form in large-$N$ CFT's and beyond}
\label{sec:qbe}

In this section, we examine the conductivity of CFTs such as the critical point of the $O(N)$ model in a perturbative
$1/N$ expansion away from the free theory obtained for $N=\infty$, with a focus on the 
emergent pole structure. We are thus approaching a general correlated CFT from the free quantum gas limit,
as illustrated in the l.h.s.\ of \rfig{big-pic},
in contrast to the holographic approach.
Our main example, though not the only one, is the $O(N)$ NL$\s$M.
%Damle and Sachdev, and Sachdev showed that the conformally invariant fixed point of the $O(N)$ model in two dimensions is 
%characterized by a finite and universal charge conductivity, both in the DC, $\w/T\ll 1$, and AC, $\w/T\gg 1$, limits. 
%Their work established the importance of taking the limits of small temperature and small frequency in the
%appropriate order to capture the DC or AC response, which are characterized by different scattering mechanisms, 
%and as such differ in general. Regarding the actual form of the conductivity as a function of $\w/T$, 
%there have been previous observations by the aformentioned authors that the small frequency part of $\s$ at leading order in $1/N$ 
% ressembles a Drude form:
%Revisiting these results, we establish that it is more than ressemblance: 
We show that the small-frequency quantum critical conductivity in the large-$N$ limit accurately satisfies the Drude form:
\begin{align}\label{eq:drude}
  \s(\w)=\frac{\s_0}{1-i\w\tau}\,.
\end{align}
The quantum Boltzmann equation (QBE) approach in the hydrodynamic regime thus captures the leading QNM at small 
frequencies, but is limited in that it misses the other poles and all the zeros. 
Although it would be desirable to have a method that captures the full analytic structure of the conductivity of CFTs
such as the $O(N)$ model, the Drude pole nonetheless contains essential information in the d.c.\ limit. 
In addition, we can use the Drude form to verify \emph{small-frequency} conductivity sum-rules. 

The fact that a single pole can capture the small-frequency complex conductivity at large but finite $N$ can
seem a priori surprising given that 
the QBE that is solved to obtain $\s$ is fairly complicated, including both elastic and
inelastic scattering of the critical quasiparticles. Below, we shed light on previous analyzes\cite{damle,sachdev-book,will} 
by providing a transparent form for the
solution to the QBE, which leads to the emergent Drude behavior of the low-frequency conductivity. 
Although we focus mainly on the $O(N)$ model, we provide similar results for a particular gauged $O(N)$
model as well as for a fermionic CFT. 

\begin{figure}
\centering
\subfigure[]{\label{fig:SI-rot}\includegraphics[scale=.55]{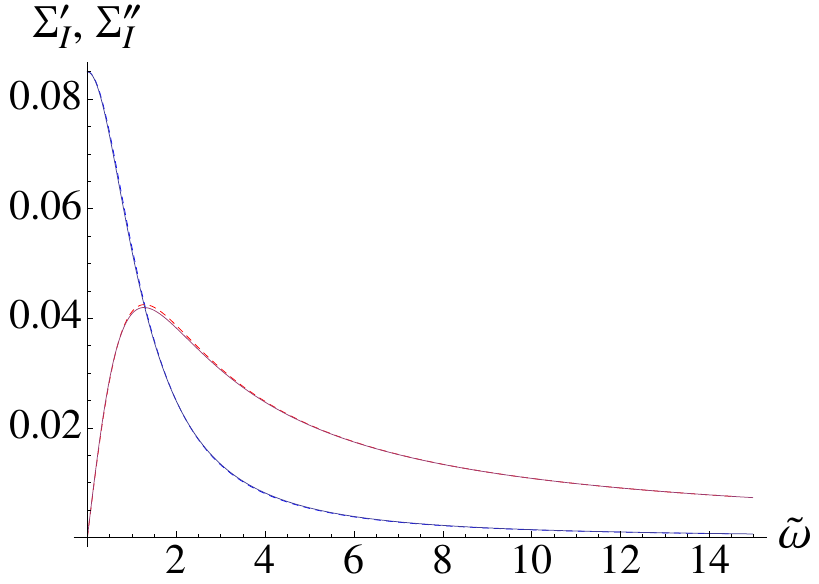}}
\subfigure[]{\label{fig:F-fn}\includegraphics[scale=.7]{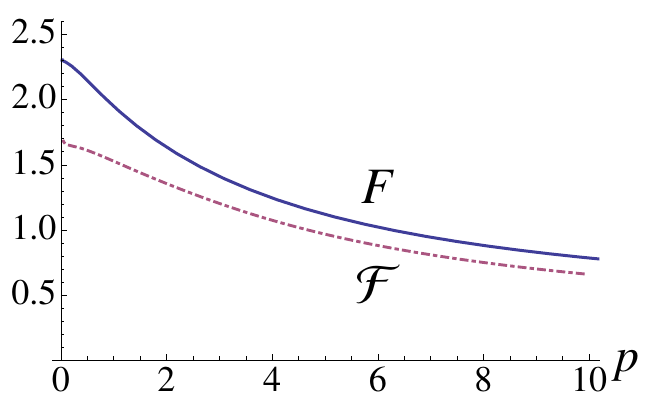}}
\caption{\label{fig:sig-drude} a) Universal scaling function for the small-frequency conductivity, $\S_I(\t\w)$,
of the quantum critical $O(N)$ model. The solid lines correspond to the numerical solution of the non-trivial QBE,
while the dashed ones to the Drude form fit. b) The momentum-dependent $F(p)$ function entering the kernel of the QBE, \req{QBE-lin},
and the renormalized $\mc F$ function determining the solution of the QBE, \req{scriptF}.
} 
\end{figure}

Let us first consider the case of the pure $O(N)$ model. We focus on the small frequency limit, $\w\ll T$, 
where the conductivity $\s$ adopts the universal scaling form\cite{damle,sachdev-book}
\begin{align}\label{eq:scaling}
  \s=\frac{e^2}{\hbar}\times N\S_I\left(\frac{N\w}{T}\right)\,,
\end{align}
where $e$ is the quantum of charge, and the subscript $I$ in the scaling function $\S$ reminds us that it is valid only at small frequencies, $\w\ll T$.
The factors of $N$ are such that the small-frequency conductivity becomes a delta function at $N=\infty$, the free limit.
For $\w \ll T$, the conductivity contains important contributions from the incoherent \emph{inelastic} scattering processes between the bosons. 
When $N$ is very large these scattering processes can be treated perturbatively in $1/N$\cite{ssqhe,sachdev-book}.
We now present the essence of the QBE approach and the results; further details can be found in 
Refs.~\onlinecite{ssqhe,sachdev-book,will}. 
Under an applied oscillatory electric field that couples to the charge, the distribution functions of the
bosonic positive/negative ($+/-$) charge excitations are modified to linear order according to
$f_\pm(\b k,\w)= n_B(\e_k)2\pi\de(\w)+s\b E\cdot \b k\vph(k,\w)$. (Note that the $O(N)$ model has many 
conserved charges, but we pick one and couple the ``electric field'' to it.)
It can be shown that the linearized
QBE for the deviation $\vph$ takes the form\cite{sachdev-book,will}:
\begin{align}\label{eq:QBE-lin}
  -i\t\w\vph+g(p)=-F(p)\vph +\int dp'K(p,p')\vph(p')\,,
\end{align}
where we have rescaled the frequency, $\t\w=N\w/T$, defined the dimensionless momentum $p=k/T$, and absorbed
factors of $T$ and $N$ into the unknown function $\vph$. The r.h.s.\ is the linearized collision term arising from the interactions
between the quantum critical modes appearing at order $1/N$.
The r.h.s.\ is the linearized collision term arising from the interactions
between the quantum critical modes appearing at order $1/N$. In the NL$\s$M formulation, the system consists of 
a vector field coupled to a single Lagrange multiplier field that enforces the unimodular constraint for the former.
The collision term arises from interactions between the vector field and the Lagrange multiplier, the latter aquiring dynamics
at order $1/N$. 
It contains two terms: the first, depending on 
a function $F$ (see \rfig{F-fn}), encodes elastic scattering processes; $F$ is essentially a momentum dependent scattering rate. 
The second term involves an integral over a kernel $K$ and it encodes the inelastic scattering processes with the Lagrange
multiplier field. 
On the l.h.s.\ the function $g(k/T)=T\pd_{\e_k}n_B(\e_k)$ acts as ``source''
for the QBE, where $\e_k^2=\De(T)^2+k^2$ and $\De\propto T$. More details regarding this
temperature dependent mass (inverse correlation length) can be found in
Appendices~\ref{app:sum} and \ref{app:ana}.
Solving the equation numerically, we find that to great precision the solution satisfies the simple form
\begin{align}\label{eq:scriptF}
  \vph(p,\t\w)=\frac{g(p)}{i\t\w-\mc F(p)}\,,
\end{align}
where $\mc F(p)$ is a monotonous function whose behavior closely resembles that of $F(p)$, \req{QBE-lin}, as can
be seen in \rfig{F-fn}.
The case $\mc F=F$ would be the exact solution in the absence of the kernel $K$ in the 
r.h.s.\ of \req{QBE-lin}. (The latter complicates the analysis and prevents analytical solubility.)
We see that the effect the kernel $K$ is to renormalize $F$ to $\mc F$, which encodes all the information
about the non-trivial inelastic scattering processes.
The corresponding solution for the conductivity is shown in \rfig{SI-rot}; it can be obtained\cite{sachdev-book,will} 
by integrating
$\vph$:
\begin{align}\label{eq:integral-sig}
  \s(\w\ll T)=\frac{e^2}{\hbar}N\times \underbrace{\frac{1}{2\pi}\int_0^{\La/T} dp\, 
    \frac{p^3\vph(p,\t\w)}{\e_p}}_{\S_I(\t\w)}\,,
\end{align}
where $\La$ is a momentum cutoff that is used in the numerical solution. We note that as $\vph$ decays
exponentially at large momenta, a cutoff can be safely used. 
Interestingly, the resulting conductivity is found to obey a Drude form to great accuracy:
\begin{align}\label{eq:cft-drude}
  \S_I(\t\w)=\frac{\S(0)}{1-i\bar\tau\t\w}\,,
\end{align}
where $\S(0)=0.085$ and $\bar\tau=\tau/T=0.775$ are two universal numbers that characterize the
\emph{entire} low-frequency charge response. The former yields the d.c.\ conductivity while the latter is a dimensionless scattering rate: 
\begin{align}\label{eq:sig0}
   \s_0&=\frac{e^2}{\hbar}\times N\S(0)\,, \\
   \tau&=\frac{N\bar\tau}{T}\,. \label{eq:tau-drude}
\end{align}
%$\s_0=N\S(0)e^2/\hbar$, 
The plot for the Drude form
is shown with dashed lines in \rfig{SI-rot}. The numerical solution and the Drude forms are nearly
indistinguishable over the entire range $0\leq \t\w <14.5$. 
% We can thus write the frequency-dependent QC conductivity in the Drude form, \req{drude}, with
The emergent scattering rate $1/\tau$ gives the location of the only pole of the conductivity in this limit:
\begin{align}\label{eq:Drude-pole-O(N)}
  \w_{\rm Drude}=-i\frac{T}{N\bar\tau}.  
\end{align}
As $N$ grows, the pole approaches the origin along the imaginary axis in the LHP; once it reaches it,
the low-frequency conductivity becomes a delta function, as shown by the arrow in \rfig{sketch-rotor}. 
The $N=\infty$ conductivity is singular and cannot be described by a meromorphic function. This is to be
expected since it describes the transport of a free gas of bosons as opposed to a generic correlated
CFT. 

Although a Drude-like low-frequency conductivity can be expected from the broadening of the zero-frequency
delta function by interactions\cite{damle}, we do not have a complete understanding regarding the excellent
quantitative agreement mentioned above. We observe that many different deviation functions $\vph$ can
give rise to a conductivity that is very well characterized by the Drude form. For example, one could use
$\vph(p)=1+1/(1+p)$ in \req{integral-sig} and obtain a very accurate Drude form. At the same time, numerous choices
would yield clear deviations. One ingredient that seems to contribute to the Drude form is the
presence of a non-parametrically small temperature dependent mass for the excitations, $\De\sim T$. 
In contrast, in the Wilson-Fisher fixed point accessed by dimensional expansion in $\varepsilon=3-d$, where
$d$ is the spatial dimension of the $O(N)$ model, the mass in the QBE can be neglected
at leading order in $\varepsilon$. The resulting conductivity does not agree as well with the single-pole form.
A further example can be found below where we consider a CFT of Dirac fermions. The QBE for the conductivity
can again be solved by ignoring the temperature-dependent mass to leading order\cite{ssqhe}, and we find that
although the Drude form fits well, it is not as a successful when compared with the large-$N$ $O(N)$ model. 
A full treatment of these questions is beyond the scope of the present paper and we leave it for future work.

At this point, we can compare these numerical results with those from the holographic analysis.
In the latter, we take $\g=1/12$, which saturates the stability bound on the particle-like side 
and should be the most appropriate to compare with the almost free large-$N$ $O(N)$ quantum critical point. 
Indeed, the further $\g$ is from the bound, the closer the effective theory is to the strongly 
interacting ``ideal quantum fluid'' limit found at $\g=0$.
At $\g=1/12$, we find that the Drude pole is located at $w_{\rm Drude}^{\rm hol}\approx -0.26i$ (see \rfig{pure-rot3} or \rfig{pure-rot5}), which
translates to $\w^{\rm hol}_{\rm Drude}=-i4\pi w_{\rm Drude}^{\rm hol}T\approx -i3.27T$. On the other hand, 
the Drude pole of the $O(2)$ model obtained by extending the result from the large-$N$ limit, \req{Drude-pole-O(N)}, 
is located at $\w_{\rm Drude}\approx -i0.65T$. The Drude pole from the QBE approach is thus located closer
to the origin compared to the one arising from the holographic analysis. We thus predict that higher $1/N$ 
corrections to the QBE will push the pole further down in the LHP. This is not surprising because the
extension of the large-$N$ result to $N=2$ yields a ratio of the d.c.\ to high frequency conductivities, $\s_0/\s_\infty$,
that is larger than within the holographic analysis:
\begin{align}
  \frac{\s_0}{\s_\infty} &= \frac{N\S(0)}{\S(\infty)}\xrightarrow{N=2} 2.13\,, \quad \textrm{large-$N$ $O(N)$ model} \\
  \frac{\s_0}{\s_\infty} &= 1+4\g= 1.33 \,, \quad \textrm{holography}
\end{align}
where we have used $\S(\infty)=(1-8\eta/3)/16\xrightarrow{N=2} 0.03998$ as the large-frequency 
scaling function for the conductivity of the $O(N)$ model
at order $1/N$, with $\eta\propto 1/N$ being the anomalous dimension of the boson field\cite{cha}. 
It is expected that higher order $1/N$ corrections will decrease this ratio and will thus push
the Drude pole further away from the origin.

\subsection{Interactions spread the weight}
Using the above quasi-exact Drude dependence, we can examine the sum rule for the 
low-frequency part of the conductivity. This is a limited version of the 
sum rules for the full universal conductivity, \req{sum-rule1} and
\req{sum-rule2}. The sum rule reads
\begin{align}\label{eq:rotor-sum-rule}
  \int_0^\infty d\t\w\, \Re \S_I(\t\w)= \pi D/4=0.1723506\dots\,,
\end{align}
where we have defined the constant
\begin{align}\label{eq:drude-D}
  \pi D=\int_\Theta^\infty dx \left(1+ \frac{\Theta^2}{x^2}\right)\frac{1}{e^x-1}=
  0.689403\dots\,,
\end{align}
where $\Theta=2\ln[(1+\sqrt{5})/2]$ is twice the natural logarithm of the golden ratio. The integral
involving the Bose-Einstein function follows simply from the expression of the conductivity in the free theory at $N=\infty$,
see Appendix~\ref{app:ana}.
In that limit, the low frequency part of the conductivity reads $\Re\s_I(\w)=(T\pi D/2)\de(\w)$. 
On the other hand the Drude form, \req{cft-drude}, satisfies the following relation: 
\begin{align}
   \int_0^\infty d\t\w\, \Re\S_I(\t\w)=\int_0^\infty d\t\w\, \Re\left\{\frac{\S(0)}{1-i\bar\tau\t\w} \right\} 
   =\frac{\pi}{2}\frac{\S(0)}{\bar\tau}=0.17221\dots
\end{align}
where in the last equality we have used the result given above for $\S(0)$ and $\bar\tau$. 
We find that the emergent Drude form satisfies the sum rule \req{rotor-sum-rule}
within a margin of $10^{-4}$, leaving plenty of room for numerical uncertainty. We thus see that the
interactions generated at order $1/N$ spread the weight of delta function over a finite Drude peak,
whose area corresponds exactly to that of the $\de$-function of the free theory at $N=\infty$. Not only
is this an excellent check on the calculation, it also provides a constraint between the 
location of the Drude pole and the value of the d.c.\ conductivity. We are effectively left with
a single universal number characterizing the small-frequency behavior of the complex conductivity at low frequencies.
\begin{figure}
\centering
\includegraphics[scale=.55]{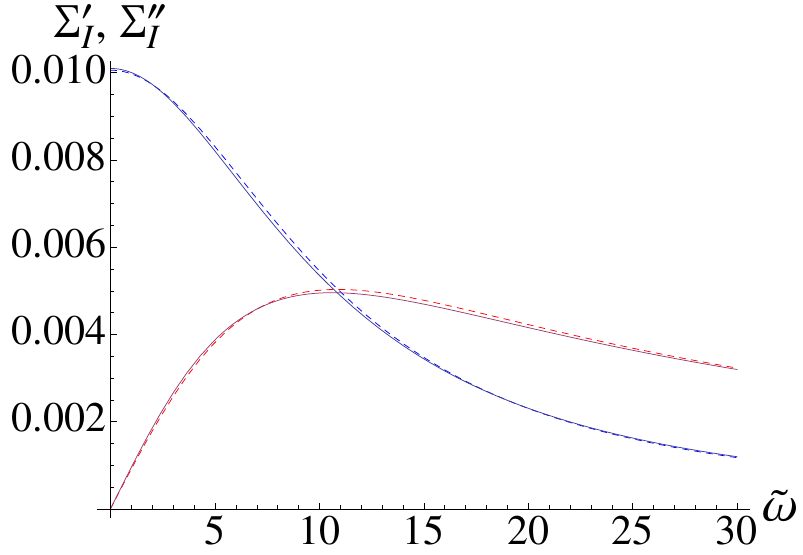}
\caption{\label{fig:sig-mit} Universal scaling function for the conductivity $\S(\t\w)$
of the gauged $O(N)$ model, with damped gauge field.
The solid lines correspond to the numerical solution of the non-trivial QBE,
while the dashed ones to the Drude form.} 
\label{fig:gauged-rot}
\end{figure}

\subsection{Flattening the conductivity with gauge bosons}
We now consider an interesting application of the above sum rule
to a gauged $O(N)$ model, where the gauge field is 
Landau damped by a Fermi surface of spinons\cite{senthil08,will}, which breaks conformal invariance of the critical point. 
This field theory was shown to be relevant to the quantum critical Mott
transition from a metal to quantum spin liquid\cite{senthil08}, as well as for the quantum critical transition between 
a N\'eel-ordered Fermi-pocket metal and a non-FL algebraic charge liquid, called a ``doublon metal''\cite{kaul}. 
It was shown\cite{will} that the same scaling form, \req{scaling}, holds as for the pure rotor model, \req{scaling}, 
since only the static gauge fluctuations contribute, the dynamical ones being strongly quenched by the Landau damping. 
This phenomenon was referred to as a ``fermionic Higgs mechanism''\cite{kaul}. The numerical solution to the
QBE including the static gauge fluctuations is shown in \rfig{gauged-rot} (for details, see Ref.~\onlinecite{will}). 
As in the case of the pure $O(N)$
CFT, it obeys a Drude form, \req{drude} with Drude parameters \req{sig0} and \req{tau-drude}, this time with
numerical values:
\begin{align}
  \S(0)=0.010\,, \qquad \bar\tau=0.092\,.
\end{align}
The d.c.\ conductivity $\S(0)$ is smaller than in the un-gauged $O(N)$ model due to the additional scattering channel:
the gauge bosons. The static gauge fluctuations are actually quite strong
and thus appreciably decrease the scattering time. The numerical
solution and the Drude form agree very well again. Note the large range of scaled frequencies over
which the agreement occurs. The deviations between the Drude and numerical solution 
seem slightly larger than in the pure rotor theory probably due to numerical uncertainties. 
The low-frequency sum rule for the conductivity, \req{rotor-sum-rule}, yields:
\begin{align}
  \frac{\pi}{2}\frac{\S(0)}{\bar\tau}=0.1720\dots\,,
\end{align}
differing from $\pi D/4$ by only $3.5\times 10^{-4}$. We see that as we add Landau damped gauge bosons
to the pure $O(N)$ model, we flatten the conductivity while keeping the emergent Drude form. 
The interactions, again, preserve the weight of the Drude peak. 

\subsection{Fermionic CFT} 
We now examine the conductivity in an interacting CFT of
Dirac fermions that arises in a model for transitions between fractional quantum
Hall and normal states\cite{ssqhe}. The field theory consists of two Dirac fermions with masses
$M_1$ and $M_2$ coupled to a Chern-Simons gauge field. The latter attaches flux tubes to each Dirac
fermion converting it to a Dirac anyon with statistical parameter $(1-\al)$, where $\al=g^2/(2\pi)$,
$g$ being the gauge coupling. The coupling $\al$ characterizes the strength of the long range
interaction between the Dirac quasiparticles mediated by the Chern-Simons field.
When $M_1,M_2>0$ the system is in a fractional quantum Hall state with
Hall conductivity $\s_{xy}=e^2q^2/(h(1-\al))$, where $qe$ is the electric charge of each Dirac quasiparticle.
The transition to an insulating state is obtained at the point where $M_1$ changes
sign while $M_2$ is taken to be large and constant. At the quantum critical point, 
the $M_1$ Dirac quasiparticles coupled to the Chern-Simons gauge field yield a finite and universal longitudinal conductivity,
whose small-frequency functional form is analogous to \req{scaling}:
\begin{align}
  \t\s_{xx}^{\rm qp}(\w)=\frac{q^2e^2}{\al^2h}\t\S_{xx}^{\rm qp}\left(\frac{\w}{\al^2T}\right)\,,
\end{align}
where $1/\al^2$ plays the same role as $N$ did in the $O(N)$ model and is taken be large.
To be more accurate, $\t\s$ is the response to the total electric field, including 
a contribution from the emergent Chern-Simons field. It can be simply related to the physical
conductivity\cite{ssqhe}. The superscript ``qp'' reminds us that this is the low-frequency
contribution arising from the scattering of thermally excited quasiparticles with each other;
it is simply a different notation for $\S_I$. 

\begin{figure}
\centering
\subfigure[]{\label{fig:sachdev}\includegraphics[scale=.34]{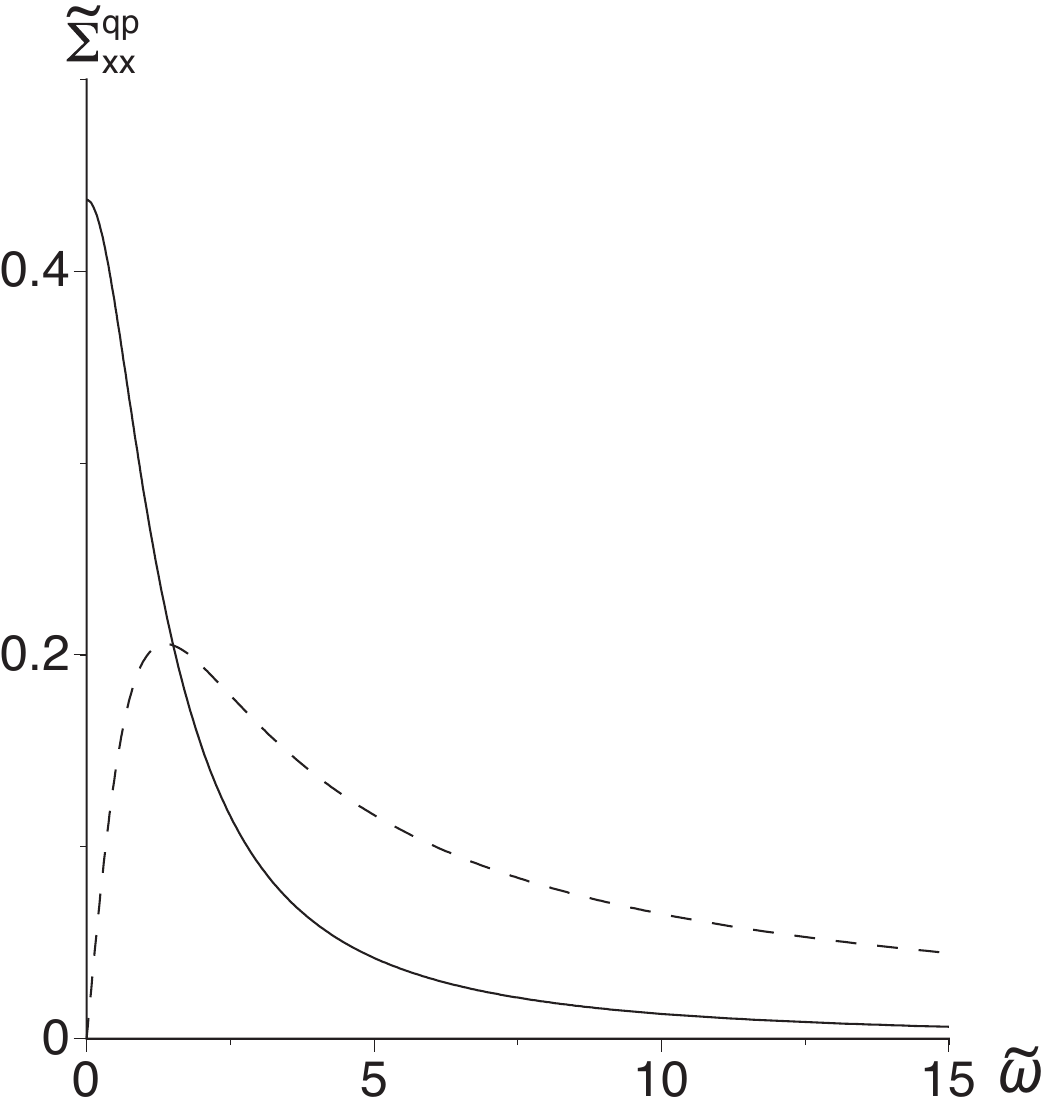}}   
\subfigure[]{\label{fig:sachdev-drude} \includegraphics[scale=.56]{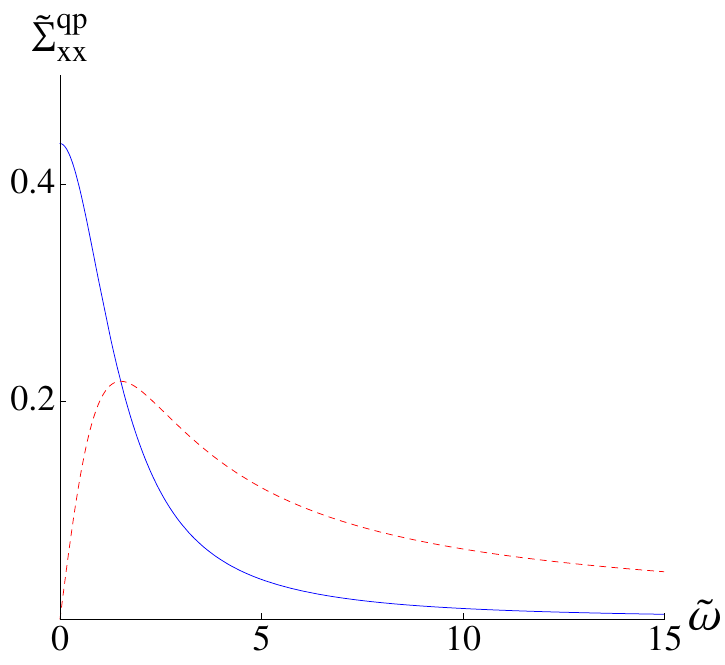}}
\caption{\label{fig:sig-fermion-cft} Universal scaling functions for the conductivity
of interacting Dirac fermions a) as computed by solving a QBE\cite{ssqhe}, 
b) from the Drude form fitted to a).}
\end{figure}

A QBE was numerically solved\cite{ssqhe} to leading order in $\al^2$, and the result is reproduced in
\rfig{sachdev}, while the corresponding Drude form fit is shown in
\rfig{sachdev-drude}. Again, both plots agree very well. The two universal Drude parameters
extracted from the fit are:
\begin{align}
  \S_{xx}^{\rm qp}(0)\approx 0.437\,, \qquad \bar\tau\approx 0.664\,.
\end{align}
The sum rule for the model is given in Ref.~\onlinecite{ssqhe}: 
\begin{align}
  \int_0^\infty \frac{d\t\w}{\pi}\Re[\t\S_{xx}^{\rm qp}(\t\w)]=\frac{\ln 2}{2}=0.3466\dots\,,
\end{align}
where $\t\w=\w/(\al^2T)$.
By using the Drude form $\S_{xx}^{\rm qp}(\t\w)=\S_{xx}^{\rm qp}(0)/(1-i\bar\tau\t\w)$, we find 
\begin{align}
  \int_0^\infty \frac{d\t\w}{\pi}\Re[\t\S_{xx}^{\rm qp}(\t\w)]\approx 0.33
\end{align}
The agreement is again quite good.

In summary, we have shown that the Drude form with its single pole captures well the low-frequency
hydrodynamic conductivity of different CFTs, a fact that was not appreciated before.
We have also seen that such a description holds for a deformation of the $O(N)$ model to 
include nearly static gauge modes. Low frequency sum rules where verified in all the models, and
serve as a useful guide in the study of interactions on the charge response.

\section{Conclusions}
\label{sec:conc}

The main thesis of this paper is that charge transport of CFTs in 2+1 dimensions is most efficiently described by a knowledge of the poles
and zeros of the conductivity in the lower-half of the complex frequency plane. Truncation to a small
number of poles and zeros gives an accurate description of the crossover from the hydrodynamic physics at small frequencies to the quantum-critical physics at high frequencies,
as was shown in Section~\ref{sec:truncate}. Such truncated forms can be used as a comparison ground with experimentally
or numerically measured charge response at conformal quantum critical points.
We also showed that the conductivity of CFTs with a global U(1) symmetry
exactly obeys two sum rules, \req{sum-rule1} and \req{sum-rule2}, for the conductivity and its (S-dual) inverse. 
The holographic computations presented here are the first to satisfy both sum rules, 
while earlier quantum Boltzmann-theory computations satisfy only one of them. 

In the holographic approach, the poles and zeros of the conductivity 
are identified with quasi-normal modes of gauge field fluctuations in the presence of a horizon.
These quasi-normal modes are the proper degrees of freedom for describing quantum critical transport, 
replacing the role played by the quasiparticles
in Boltzmann transport theory. We presented results for the quasi-normal mode frequencies in an effective holographic theory for 
CFTs which kept up to four derivative terms in a gradient expansion.

We expect that the quasi-normal modes will help describe a wide variety of dynamical phenomena in strongly-interacting 
quantum systems, including those associated with deviations from equilibrium \cite{bhaseen}. 
The quasi-normal mode poles and zeros should also help in 
the analytic continuation of imaginary time data obtained from quantum Monte Carlo simulations.

\acknowledgements
We are grateful for many enlightening discussions with B.~Burrington, S.-S.~Lee, A.~Singh,
and X.-G.~Wen. We also wish to thank P.~Ghaemi, A.~G.~Green, M.~Killi, Y.-B.~Kim, J.~Maldacena, J.~Rau, T.~Senthil
and R.~Sorkin for useful conversations. 
This research was supported by the National Science Foundation under grant DMR-1103860 and by 
the Army Research Office Award W911NF-12-1-0227 (SS), as well as by a Walter Sumner Fellowship (WWK).
SS acknowledges the hospitality of the Perimeter Institute, where significant portions of this work
were done. Research at Perimeter Institute is supported by the Government of Canada through Industry Canada and by the Province of Ontario through the Ministry of Research \& Innovation.
\appendix 

\section{Conductivity sum rules}
\label{app:sum}

Conductivity sum rules are familiar in condensed matter physics in systems with a finite lattice cutoff. 
The standard derivation starting from the Kubo formula for a general Hamiltonian, $\mathcal{H}$, yields \cite{mahan}
\beq
\mathcal{I} \equiv \int_0^\infty d \omega \, \Re \, \sigma (\omega ) = -\frac{\pi}{2} \lim_{\b q \rightarrow 0} \, \frac{1}{ q^2 V} \left\langle
\left[ \left[ \mathcal{H}, \rho (\b q) \right], \rho (-\b q) \right] \right\rangle,
\label{eq:sum1}
\eeq
where $\rho (\b q)$ is the density operator at wavevector $\b q$, and $V$ is the system's volume.
It is now our task to understand the structure of the commutators on the right-hand-side in the scaling limit appropriate for a 
CFT in 2+1 dimensions.

In quantum field theory, the r.h.s.\ of \req{sum1} has the structure of an ultraviolet divergent Schwinger contact term \cite{schwinger}. 
The divergence is acceptable to us, because the sum rule in \req{sum-rule1} is convergent only after the subtraction of the 
constant $\sigma_\infty$ term. The important issue for us is whether the r.h.s. of \req{sum1} has any finite corrections which depend upon
infrared energy scales such as the temperature or chemical potential ($\mu$). If such finite corrections are absent, then the sum rules in 
 \req{sum-rule1} and \req{sum-rule2} follow immediately, because $\sigma_\infty$ is the value of the $\sigma (\omega)$
at $T=0$ and $\mu =0$, and the integral is independent of $T$ and $\mu$.

It is useful to analyze this issue first for a simple CFT of free Dirac fermions. 
Here we can regularize the Dirac fermions on a honeycomb lattice (as in graphene).
Fortunately, such a sum rule analysis for the honeycomb lattice has already been carried out in Ref.~\onlinecite{carbotte}. 
On a lattice with spacing $a$, Fermi velocity $v_F$, temperature $T$, and chemical potential $\mu$, they find when $T$ and $\mu$
are smaller than the bandwidth that
\beq
\mathcal{I} = c_1 \frac{v_F}{a} + \frac{a^2 T^3}{v_F^2} f(\mu/T), \label{eq:graphene}
\eeq
for some constant $c_1$ and function $f$.  
Observe that this is divergent in the continuum limit
($a \rightarrow 0$ at fixed $v_F$, $T$, $\mu$), but
the leading portion dependent upon $T$ and $\mu$ 
vanishes. So there is no dependence of $\mathcal{I}$ of the CFT upon $\mu$ and $T$.

Let us now carry out the corresponding analysis for the large-$N$ limit of the $O(N)$ rotor model. This is an interacting 
theory at finite $N$, and we will
see that the scaling limit has to be taken carefully so that we remain properly in the vicinity of the conformal fixed point in the presence
of infrared perturbations like $T$ or deviations from the critical point.
We regularize the rotor model on a square lattice
of sites $i$,$j$, spacing $a$, with the Hamiltonian
\beq
\mathcal{H} =   \frac{g a^2}{2 N} 
\sum_i \hat{\pi}_{i\alpha}^2 + 
\frac{c^2 N}{2g} \sum_{\langle ij \rangle} (\hat{\phi}_{i\alpha} - \hat{\phi}_{j\alpha} )^2 ,
\label{eq:rotorON}
\eeq
where $\hat{\phi}_{i \alpha}$, with $\alpha = 1 \ldots N$ are the rotor co-ordinates which obey the constraint
\beq
\sum_{\alpha} \hat{\phi}_{i \alpha}^2 = 1 \label{eq:sum2}
\eeq
at all sites $i$. The $\hat{\pi}_{i \alpha}$
are their conjugate momenta with
\beq
[\hat{\phi}_{i\alpha} , \hat{\pi}_{j\beta}] = i  \delta_{\alpha\beta} \frac{\delta_{ij}}{a^2}.
\eeq
The coupling constant $g$ is used to fix the model in the vicinity of the critical point at $g=g_c$, and we will
take the continuum limit $a \rightarrow 0$ at fixed velocity $c$ and $T$. In the large $N$ limit, the critical point is at
\beq
\frac{1}{g_c} = \int_{\b k \in \text{BZ}} \int \frac{d \omega}{2 \pi} \frac{1}{\left[ \omega^2 + 2 (c/a)^2 (2 - \cos (k_x a) - \cos (k_y a)) \right] };
\eeq
This determines $g_c \approx 3.11 a  c$.
If we move away from the critical point, or to non-zero temperatures, 
then the Lagrange multiplier enforcing the constraint \req{sum2} induces an energy gap $\Delta (T)$
determined by
\beq
\frac{1}{g} = \int_{\b k \in \text{BZ}} T \sum_{\omega_n} \frac{1}{\left[ \omega_n^2 + 2 (c/a)^2 (2 - \cos (k_x a) - \cos (k_y a)) + \Delta^2 (T) \right] }\,, \label{eq:sum4}
\eeq
where $\w_n$ are the bosonic Matsubara frequencies.
We will take the limit $a \rightarrow 0$ at fixed $\Delta (T)$ and $T$. In this limit we have
\beq
\frac{1}{g} = \frac{1}{g_c} - \frac{\Delta (0)}{4 \pi}.
\eeq

The density operator is 
\beq
\rho (\b q) = a^2 \sum_{i} e^{- i \b q \cdot \b r_i} l_{\alpha\beta} \, \hat{\phi}_{i\alpha} \,  \hat{\pi}_{i\beta},
\eeq
where $l_{\alpha\beta}$ is one of the antisymmetric generators of $O(N)$ normalized so that $\mbox{Tr}(l^2) = -1$.
Evaluating the commutator in \req{sum1} we find
\beq
\left[ \left[ \mathcal{H}, \rho (\b q) \right], \rho (-\b q) \right] = -\frac{2 c^2}{g} \sum_{\langle ij \rangle} \hat{\phi}_{i\alpha} 
\hat{\phi}_{j \alpha} | e^{i \b q \cdot \b r_i} -  e^{i \b q \cdot \b r_j} |^2\,.
\eeq
So taking the limit the long wavelength limit yields
\beq
\lim_{\b q \rightarrow 0} \, \frac{1}{ q^2} 
\left[ \left[ \mathcal{H}, \rho (\b q) \right], \rho (-\b q) \right] = - \frac{c^2 a^2}{g} \sum_{\langle ij \rangle}  \hat{\phi}_{i\alpha} 
\hat{\phi}_{j \alpha}\,.  \label{eq:sum5}
\eeq
Using \req{sum2}, we can now write the conductivity sum rule as
\Beq
\mathcal{I} &=& \frac{\pi c^2}{2 g} - \frac{\pi c^2 a^2}{4 g V} \sum_{\langle ij \rangle} \left\langle (\hat{\phi}_{i\alpha} - \hat{\phi}_{j\alpha} )^2 \right \rangle \nonumber \\
&=& \frac{\pi c^2}{2 g} - \frac{\pi c^2}{2} \int_{\b k \in \text{BZ}} T \sum_{\omega_n}  \frac{(2 - \cos (k_x a) - \cos (k_y a))}{\left[ \omega_n^2 + 2 (c/a)^2 (2 - \cos (k_x a) - \cos (k_y a)) + \Delta^2 \right] }.
\Eeq
Evaluating the frequency summation, and then taking the limit $a \rightarrow 0$, we obtain the expansion
\beq
\mathcal{I} = \frac{\pi c^2}{2 g} - \al_1 \frac{c}{a} + \al_2 \frac{\Delta^2}{c} a - a^2 \frac{\pi c^2}{4} \int_0^{\infty} \frac{d^2 k}{4 \pi^2} \frac{k^2}{\sqrt{c^2 k^2 + \Delta^2} (e^{\sqrt{c^2k^2 + \Delta^2}/T} -1)} + \ldots \label{eq:sum6}
\eeq
where $\al_1\approx 0.75$ and $\al_2\approx 0.13$.
The crucial feature of this result is that there is no term $\sim \Delta$, and all terms containing $\Delta$ vanish as $a \rightarrow 0$.
A term $\sim \Delta$ does appear if we choose a general $\Delta$ which does not obey \req{sum4} and then evaluate \req{sum5}.
Thus
the imposition of the constraint \req{sum2} at all $T$ was important for the absence of such a term. The general features of 
\req{sum6} are similar to \req{graphene}, and so the same conclusions apply.

\section{Analytic structure in the $N \rightarrow \infty$ limit of the $O(N)$ model}
\label{app:ana}

This appendix notes a few features of the conductivity of the $O(N)$ rotor model in the complex frequency plane, in the $N \rightarrow \infty$ limit.
For the model in \req{rotorON}, the conductivity as a function of the complex frequency $z$ 
follows from Ref.~\onlinecite{damle}:
\beq
\sigma (z) = \frac{i TD}{z} + 
\frac{iz}{4 \pi} \int_{\Delta}^{\infty} d \Omega \frac{(\Omega^2- \Delta^2)}{\Omega^2 (z^2 - 4 \Omega^2)} \coth \left(\frac{\Omega}{2 T} \right)\,,
\label{eq:sigmaz}
\eeq
where the contour of $\Omega$ integration determines the specific choice of the current correlator,
and the Drude weight scales linearly with the temperature. We have defined the numerical constant
\beq
D = \frac{1}{8 \pi} \int_{\Delta}^\infty d \Omega \frac{(\Omega^2- \Delta^2)/T^2}{\Omega \sinh^2 (\Omega/(2T))}.
\eeq
whose value is given in \req{drude-D}.

The retarded response function $\sigma^R (z)$ is obtained by choosing $z$ in the UHP, and the contour of integration
along the real frequency axis. This function $\sigma^R (z)$ is analytic in the UHP, and has a pole at $z=0$ and 
branch points at $z = \pm 2 \Delta$. 
We can perform the analytic continuation of $\sigma^R (z)$ into the lower-half plane by deforming the contour
of $\Omega$ integration into the lower-half plane, so that it is always below the points $ \pm z/2$.
Because of the presence of these branch points, the analytic continuation of $\sigma^R(z)$ into the lower-half
plane is not unique, and depends upon the path of $z$ around the branch points. 
This is a key difference from the holographic results of the present
paper, which had no branch points and a unique analytic continuation into the LHP. We expect that fully incorporating $1/N$ corrections
will make the $O(N)$ model result similar to the holographic computation. We have already demonstrated this for the case of the pole at $z=0$,
which becomes a LHP Drude pole. However a careful analysis of $1/N$ corrections determining the fate of the branch points
at $z = \pm 2 \Delta$ has not yet been carried out.

In any case, the physical value on the real axis $\sigma^R (\omega + i 0^+)$ is unique,
and was shown in \rfig{sig-full-O(N)}.
 At the critical point, this is to be evaluated at
$\Delta = \Theta T$, where $\Theta = 2 \ln ((\sqrt{5}+1)/2)$. Curiously, for this value of $\Delta$, we find zeros of the conductivity
on the real axis branch points, with $\sigma^R (\pm 2 \Theta T + i 0^+) = 0$.
So the structure of poles and zeros of the $N=\infty$ conductivity has a remarkable similarity to the $\gamma > 0$ holographic results,
as was reviewed in \rfig{summary}.
The pole at $z=0$ of the $N=\infty$ theory corresponds to the closest pole on the negative imaginary axis
of the holographic result, as we have already noted. And the zeros at $z = \pm 2 \Theta T$ of theory correspond to the two zeros closest to the real
axis in Fig.~\ref{fig:shp}.

Finally, we can verify that the sum rule in \req{sum-rule1} is satisfied by \req{sigmaz}
\beq
\int_0^{\infty} d \omega \left[ \Re \, \sigma^R ( \omega + i 0^+) - \frac{1}{16} \right] = 0\,.
\eeq
where we have used $\s_\infty=1/16$.
Note that this result is obeyed {\em only} for $\Delta = \Theta T$, and not for other values of $\Delta$, as is 
expected from the considerations
in Appendix~\ref{app:sum}. Also, as noted in the introduction, the inverse sum rule in \req{sum-rule2} 
is not satisfied by \req{sigmaz}. Although $\sigma(\omega)$ has a zero at $\omega = 2\Delta$, the location of the branch point,
this nevertheless leads to an integrable divergence in $\Re[1/\sigma(\omega)]$ at that point. 
We have indeed verified that the integral of $\Re[1/\sigma(\omega)]-\sigma_\infty^{-1}$ is finite (actually,
it is greater than unity), proving that the conductivity of the critical 
$O(N\rightarrow \infty)$ model does not respect the S-dual sum rule.

Let us also mention that the analytic structure of response functions of 
the $O(N)$ model was also examined recently 
in Ref.~\onlinecite{podolsky} away from the CFT critical point, but at $T=0$.
In the ordered phase with broken $O(N)$ symmetry, poles were found in the lower-half
plane corresponding to the Higgs excitations damped by multiple spin-wave emission.

\section{Differential equation for the numerical solution of the conductivity}
\label{ap:solv-sig}
\begin{figure}
\centering%
\subfigure[$\Re\{ \s(\g=1/12)\}$]{\label{fig:pure-rot5}\includegraphics[scale=.36]{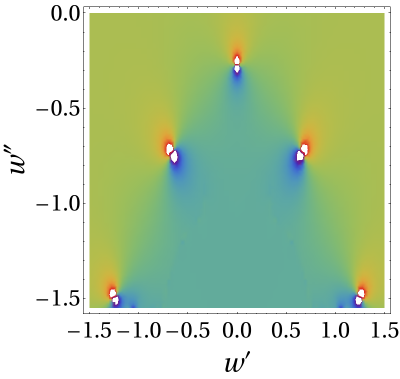}}  
\subfigure[$\Re \{\hat\s (\g=1/12)\}$]{\label{fig:gauged-rot5} \includegraphics[scale=.36]{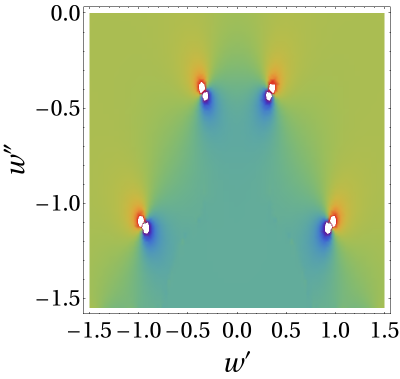}}\\
\subfigure[$\Re\{\s(\g=-1/12)\}$]{\label{fig:gauged-rot6} \includegraphics[scale=.36]{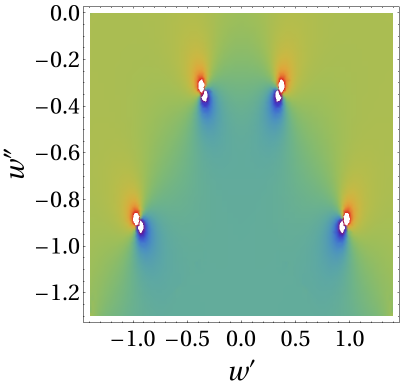}}
\subfigure[$\Re\{\hat \s(\g=-1/12)\}$]{\label{fig:pure-rot6}\includegraphics[scale=.36]{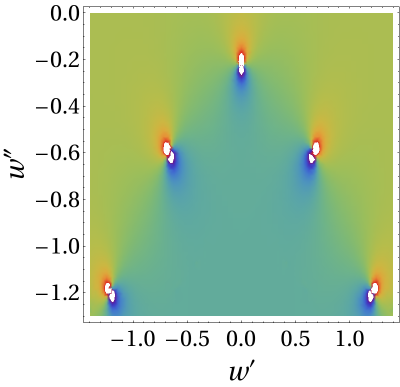}}  
\caption{\label{fig:sig2} Conductivity $\s$ and its dual $\hat\s=1/\s$ in the LHP, $w''=\Im w \leq 0$, for
$|\g|=1/12$. There is a qualitative correspondence of the pole structure between $\s(w;\g)$ and
$\hat\s(w;-\g)$. Note that the poles of $\hat\s(w;\g)$ are the zeros of $\s(w;\g)$. 
} 
\end{figure}

We first factor out the singular part of $A_y$ near the horizon: $A_y=(1-u)^{-iw}F(u)$.
Making this substitution in the EoM for $A_y$, \req{Ay-ode}, we obtain the following differential
equation for $F$:
\begin{multline}
0= F''- \left( \frac{3 u^2 (1-4 (1-2 u^3) \gamma )}{(1-u^3) (1+4 u^3 \gamma )}-
\frac{2 i w }{1-u} \right) F'\\
 + \frac{i w \left(  (1+u+u^2) (1+2 u +4u^2(3+4u+5 u^2) \gamma )-i (2+u)
 (4+u+u^2) (1+4 u^3 \gamma) w \right)}{(1-u) (1+u+u^2)^2 (1+4 u^3 \gamma )}F\,.
\end{multline}
This is to be compared with the simpler form of the equation for the full $A_y$, \req{Ay-ode}.
The two boundary conditions at the horizon read
\begin{align}
  F(1)&=1\,, \\
  F'(1)&=  \frac{i w ( i+2w +8 \g (2i+w ))}{(1+4 \g ) ( i+2w )}\,. \label{eq:bcFp}
\end{align}
The second condition follows from the solution of the differential equation near $u=1$: $F(u)\approx 1-(1-u)\digamma$,
with $\digamma$ being the r.h.s.\ of \req{bcFp}. The numerical solution is shown in \rfig{sig-direct} and in \rfig{sig2},
where the poles and zeros in the LHP can be seen more precisely.

\section{WKB analysis for asymptotic quasi-normal modes}
\label{app:wkb}
The goal of the WKB analysis is to identify the QNMs of the gauge field at large frequencies, $|w|\gg 1$.
According to the AdS/CFT correspondence, these frequencies can then be put in correspondence with the 
poles of the gauge correlation function $\mc G_{yy}$ proportional to the conductivity, \req{sig-ads-cft}. 
The standard analysis examines the
solutions to \req{sform} near 1) the black hole singularity, 2) the event horizon, and 3) the asymptotic boundary.
Matching of the solutions usually gives an expression for a set of discrete QNM frequencies. 
Generically one obtains two solution for $A_y$, with one vanishing as the boundary is approached. Discarding 
the non-vanishing one leads to a ``quantization'' condition on the QNMs.

As mentioned in the main text, the EoM for the $y$-component of the gauge field reads:
\begin{align}\label{eq:ay}
 0 &= A_y'' + \frac{h'}{h} A_y' +\frac{9w^2}{f^2}A_y  \\
  \frac{h'}{h}&= \frac{f'}{f}+\frac{g'}{g}
\end{align}
The second equality follows from $h=fg$. 
We can change coordinates to bring this equation into a Schr\"odinger form, which will
be more convenient for the analysis of the QNMs. To do so, we want to 
transform away the linear-derivative term. One way 
involves changing variables to $dx=du/f$, as we illustrate below. 

Before going into the WKB analysis, let us first review the simplest scenario, $\g=0$, 
i.e. in the absence of the function $g$ arising from the Weyl curvature coupling.
The exact solution is obtained by using the new (complex) coordinate $z$:
\begin{align}\label{eq:dzdu}
  \frac{dz}{du}:=\frac{3}{f}= \frac{3}{1-u^3}\,.
\end{align}
This puts \req{ay} in the form:
\begin{align}\label{eq:harmonic}
  \pd_z^2A_y + w^2 A_y=0
\end{align}
with solutions: $e^{\pm i w z}$. To apply the boundary condition we need to examine the
explicit form of $z(u)$. Integrating \req{dzdu}, we obtain
\begin{align}\label{eq:zu}
  z(u) = \sum_{p=1}^3 \frac{3}{f'(u_p)}\ln(1-u/u_p)
\end{align}
where $u_p$ are the 3 zeros of $f$. They are simply the cubic roots of unity: $u_p^3=1$, i.e.
\begin{align}
  u_1 &= 1 \\
  u_2 &= -(1+i\sqrt 3)/2\,, \qquad  u_3 =u_2^*
\end{align}
which is trivially found by noting that $1-u^3=(1-u)(u^2+u+1)$. We give a few properties of the
generating polynomial $f$ and its roots that will be useful for future analysis. First, the derivative
of $f$ permutes $u_2$ and $u_3$ while leaving $u_1$ invariant (up to signs): $f'(u_1)/3=-u_1$ and $f'(u_2)/3=-u_3$.
As a result, we get the following identities:
\begin{align}
  \sum_{p=1}^3 u_p &=0 \\
  \sum_{p=1}^3 \frac{u_p^n}{f'(u_p)} &= 
  \begin{cases}
   -1 & \text{if  } n \bmod 3 =2 \\
   0  & \text{otherwise}
  \end{cases}
\end{align}
Recall that we need to apply an infalling boundary condition, $A_y\approx (1-u)^{-iw}$,
near the event horizon, $u=1$. Using \req{zu}, we find that as $u\rightarrow 1$,
\begin{align}
  e^{\pm iwz}\rightarrow C_\pm \times (1-u)^{\mp iw}
\end{align}
where $C_\pm=e^{\pm i w(\ln 3+\pi/\sqrt 3)/2}$. Hence, the boundary condition selects $A_y=e^{iwz}$.
This in turn yields:
$\s=-i\frac{\pd_uA_y}{3wA_y}\big|_{u\rightarrow 0}=-i\frac{3iw}{3w(1-u^3)}\big|_{u\rightarrow 0}=1$.
As expected the conductivity of the CFT holographically dual to the Einstein-Maxwell theory on S-AdS$_4$
is constant for all complex frequencies, hence self-dual. We now include a finite $\g$, which
prevents analytical solubility, just like the $1/N$ collision term did for the $O(N)$ model.

We wish to transform \req{ay} into a Schr\"odinger form. To facilitate comparison with the literature,
notably with Ref.~\onlinecite{cardoso} which serves as a guide for our analysis, 
we shall perform the WKB analysis starting with the coordinate $r=1/u$ instead of $u$. This is the radial
holographic coordinate introduced in the main body, with the difference that it is rescaled by $r_0$.
We define $\df=r^2 f=r^2-r\inv$, and the corresponding new tortoise coordinate (the analogue of $z$ introduced
above):
\begin{align}
  \frac{dx}{dr}=\frac{1}{\df}\,.
\end{align}
In terms of $x$, the EoM for $A_y$ becomes:
\begin{align}
  \frac{d^2A_y}{dx^2}+\frac{1}{g}\frac{dg}{dx}\frac{dA}{dx}+\nu^2 A_y=0\,,\qquad \nu=3w\,.
\end{align}
We have defined the rescaled frequency $\nu$ to simplify the comparison with previous works. 
We note that in the limit where $\g=0$, the linear derivative term vanishes and we are left with
a trivial harmonic equation as above. For finite $\g$, we can remove such a term by 
%The crux of the first transformation method is given in Ref.~\onlinecite{myers11}; here, we provide extra details
%relevant for the WKB analysis, which was not done previously. 
introducing two functions to parameterize $A_y$: 
\begin{align}
  A_y=G(x)\psi(x)\,,
\end{align}
where in order for $\psi$ to satisfy an equation of the Schr\"odinger form, $G$ needs to satisfy the first
order differential equation:
\begin{align}
  \frac{dG}{dx} +\frac{1}{2g}\frac{dg}{dx}G=0\,.
\end{align}
This can be solved in general by $G=1/\sqrt g=1/\sqrt{1+4\g u^3}$.
%When $\g>-1/4$, this function is regular for $r\in[0,1]$. 
The resulting ``Schr\"odinger'' equation for $\psi$ is:
\begin{align}\label{eq:sform}
  -\frac{d^2\psi}{dx^2}+W(x)\psi=\nu^2 \psi\,,
%  -\pd_z^2\psi +W\psi=w^2\psi
\end{align}
where
\begin{align}\label{eq:W}
  W=\frac{6\g(r^3-1)}{r^4(r^3+4\g)^2}[2r^6+(2\g-5)r^3-14\g]\,.
 % W=2\g u(1-u^3)\,\frac{2-(5-2\g)u^3-14\g u^6}{3(1+4\g u^3)^2}
\end{align}
The potential $W$ prevents the exact solubility of the equation, and as expected vanishes
as $\g\ra 0$. In that limit, $G\ra 1$ and $W\ra 0$, and the equation reduces to the harmonic
one \req{harmonic}. Note that the potential vanishes at the boundary, $r=\infty$, just as the Weyl
curvature does.

\begin{figure}
\centering
\includegraphics[scale=.45]{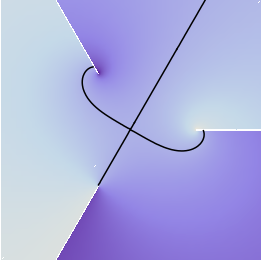}
\caption{\label{fig:stokes} The Stokes line, $\Im(\nu x)=0$, in black in the complex $r$-plane; 
$r=0$ corresponds to the intersection
point of the two branches of the Stokes line. The color shading represents the value of $\Im(\nu x)$.
The 3 branch cuts coming from the logarithms are clearly visible.
}
\end{figure}

The underlying idea of the WKB method is to examine the behavior of $A_y$ or $\psi$ on the
Stokes line in the complex $r$-plane defined via:
\begin{align}
  \Im (\nu x)=0\,.
\end{align}
The first step is thus to identify this Stokes line by studying the behavior of the
tortoise in terms of $r$.
As above, the defining relation for the tortoise can be integrated to give:
\begin{align}
  x(r)&=\frac{1}{3}\sum_{p=1}^3\frac{1}{\df'(r_p)}\ln(1-r/r_p)\\
  &=\frac{1}{3}[\ln(1-r)+\al^*\ln(1-\al^*r)+\al\ln(1-\al r)]
\end{align}
where $r_1=1, r_2=\al,r_3=\al^*=\al^2$ are the three cubic roots of unity, with $\al=(-1+i\sqrt 3)/2$;
precisely the $u_p$ introduced above.
Near $r=0,\infty$, the tortoise scales like
\begin{align}
  x&\approx -\frac{r^2}{2}\,, \quad r\ra 0 \\
  x&\approx x_0-\frac{1}{r}\,, \quad r\ra \infty \label{eq:x-infty}\\
\end{align}
respectively, where we have introduced
\begin{align}
  x_0\equiv x(r\ra \infty)=\frac{2\pi\sqrt{3}}{9}e^{-i\pi/3}\,,
\end{align}
which will play a central role in the WKB analysis\cite{cardoso}. Its value is well-defined due to the absence of monodromy
at infinity, even in the presence of the 3 branch cuts coming from the logarithms, see \rfig{stokes}. The value of $x_0$ 
dictates that of $\nu$ via $\nu x_0\in\mathbb R$: $\nu=\zeta e^{i\pi/3}$, where $\zeta\in\mathbb R$. In particular, from this
and \req{x-infty}, we see that the branch of the Stokes line that extends to infinity follows the
line $r=\rho e^{i\pi/3}$, where $\rho$ is real. Near the origin, we have $\Im(e^{i\pi/3}x)\approx-\Im(e^{i\pi/3}r^2)/2$,
which implies $r=\rho e^{-i\pi/6}$, $\rho\in\mathbb R$, in addition to $r=\rho e^{i\pi/3}$. 
These two branches of the Stokes line cross at the origin as we show in \rfig{stokes}. 
We now proceed to the WKB analysis by examining the solution to \req{sform} in the vicinity
of $r=\infty,0,1$.

{\bf Near $r=\infty$}: The potential $W(r)$ is irrelevant since $W\sim 1/r$. This is not surprising since
we expect $\g$ to be irrelevant near the UV boundary and $W\propto \g$. The equation becomes harmonic.
We write the solution in terms of the shifted variable, $x-x_0$, and use Bessel functions although simple
sines and cosines would suffice; this allows us to compare with other QNM analyses\cite{cardoso}. We have
\begin{align}
  \psi(x)=B_+\sqrt{2\pi\nu (x-x_0)}J_{j_\infty/2}(\nu(x-x_0))+B_- \sqrt{2\pi\nu(x-x_0)}J_{-j_\infty/2}(\nu(x-x_0))
\end{align}
where $j_\infty=1$, and $J_{1/2}(z)=\sqrt{2/\pi}\sin(z)/\sqrt{z}$, $J_{-1/2}(z)=\sqrt{2/\pi}\cos(z)/\sqrt{z}$.
As we have discussed in the main text, we need to impose the vanishing of $A_y=\psi G$ at the boundary, which
leads to $\psi(x_0)=0$ since $G(x_0)=1$. We thus have our first constraint, $B_-=0$.

{\bf Near $r=0$}: Near the black hole singularity, the potential diverges:
\begin{align}
  W(r)=\frac{21}{4r^4}=\frac{21/4}{4x^2}=\frac{j_0^2-1}{4x^2}\,,
\end{align}
with $j_0=5/2$. In the second inequality we have used $x\approx -r^2/2$ near the singularity. We thus have
the Bessel solution:
\begin{align}
  \psi(x)=A_+\sqrt{2\pi\nu x}J_{j_0/2}(\nu x)+A_-\sqrt{2\pi\nu x}J_{-j_0/2}(\nu x)
\end{align}

We can match the solutions near $r=\infty$ and $r=0$ using the asymptotic expansion for $z\gg 1$:
$J_a(z)\approx \sqrt{2/(\pi z)}\cos[z-(1+2a)\pi/4]$. Extending starting near the origin $r=0$,
\begin{align}
  \psi(x)&\approx 2A_+\cos(\nu x-\al_+)+2A_-\cos(\nu x-\al_-)\\
  &=(A_+e^{-i\al_+}+A_-e^{-i\al_-})e^{i\nu x}+(A_+ e^{i\al_+}+A_-e^{i\al_-})e^{-i\nu x}
\end{align}
where we have defined $\al_{\pm}=(1\pm j_0)\pi/4$. On the other hand extending from $r=\infty$ we
get 
\begin{align}
  \psi &\approx 2B_+\cos[\nu(x-x_0)-\be_+] \\
  &= B_+e^{-i\be_+}e^{i\nu(x-x_0)}+B_+e^{i\be_+}e^{-i\nu(x-x_0)}
\end{align}
where $\be_+=\pi/2$. Matching both solutions by equating the ratios of the coefficients 
of $e^{\pm i\nu x}$ yields another constraint:
\begin{align}\label{eq:wkb-constraint1}
  A_+\sin(\nu x_0+\be_+-\al_+)+A_-\sin(\nu x_0+\be_+-\al_-)=0\,.
\end{align}

{\bf Near $r=1$}: We then want to match the behavior on the Stokes branch $r=\rho e^{i\pi/3}$ with that near the
black hole event horizon $r=1$. First, we have the small-$z$ expansion $J_a(z)\approx z^aw(z)$,
where $w(z)$ is an even and holomorphic function, $w(z)= \,_0F_1(a+1;-z^2/4)/(2^a\G(a+1))$, where
$_0F_1$ is an instance of the hypergeometric function. We will rotate from the branch $r=\rho e^{i\pi/3}$,
$\rho\in\mathbb R^-$ to $r=\rho e^{-i\pi/6}$, $\rho\in\mathbb R^+$. Using $x\sim r^2$ near $r=0$, the
$\pi/2$ $r$-rotation becomes a $\pi$ $x$-rotation:
\begin{align}
  \sqrt{2\pi e^{-i\pi}\nu x}J_{\pm j_0/2}(e^{-i\pi}\nu x) &=  e^{-i(1\pm j_0)\pi/2} \sqrt{2\pi \nu x}J_{\pm j_0/2}(\nu x)\\
  &\ra 2e^{-i2\al_\pm}\cos(\nu x-\al_\pm)
\end{align}
Using this we have the following behavior on the $r=\rho e^{-i\pi/6}$, $\rho\in\mathbb R^+$ branch:
\begin{align}
  \psi(x) &\sim 2A_+e^{-i2\al_+}\cos(-\nu x-\al_+)+2A_-e^{-i2\al_-}\cos(-\nu x-\al_-) \\
  &= (A_+e^{-i\al_+}+A_-e^{-i\al_-})e^{i\nu x}+ (A_+e^{-i3\al_+}+A_-e^{-i3\al_-})e^{-i\nu x}
\end{align}
We know that at the horizon, $\psi(x)\sim e^{i\nu x}$ in order to satisfy the infalling condition,
consequently
\begin{align}\label{eq:wkb-constraint2}
  A_+e^{-i3\al_+}+A_-e^{-i3\al_-}=0\,.
\end{align}

Combining \req{wkb-constraint1} and \req{wkb-constraint2}, we find get a condition that the homogeneous
system of equations needs to satisfy in order to have a solution:
\begin{align}\label{eq:wkb-det}
  \det\begin{pmatrix}e^{-i3\al_+} & e^{-i3\al_-} \\ 
    \sin(\nu x_0+\be_+-\al_+) & \sin(\nu x_0+\be_+-\al_-)\end{pmatrix}=0
\end{align}
This equation leads to the general solution for the asymptotic QNMs:
\begin{align}
  3w x_0=\xi - 2\pi n\,, \quad n\in\mathbb N \;\;\&\;\; n\gg 1
\end{align}
where we have switched back to $w=\nu/3$.
We find two solutions for the offset parameter $\xi$:
\begin{align}
  \xi_1 &=2 i \tanh ^{-1}\left(\frac{\sqrt[4]{2}+(1+i)}{\sqrt[4]{2}+(-1-i)}\right)\approx -2.356-i 0.173\,,\\
  \xi_2 &=2 \tan ^{-1}\left(\frac{i\sqrt[4]{2}+(1-i)}{\sqrt[4]{2}+(1+i)}\right)\approx 0.785-i 0.173
\end{align}
The offset and gap, defined via $w={\rm [gap]}-n{\rm [offset]}$ for large $n$, are given by 
\begin{align}
  \rm{offset} &= \frac{\xi}{3x_0}\,, \\
  \rm{gap} &= \frac{2\pi}{3x_0} = \sqrt 3 e^{i\pi/3}\,,
\end{align}
where the offset obtained using $\xi_{1,2}$ is $-0.283 - i0.586$ or $0.150 + i0.164$, respectively.
Interestingly, we note that these results for the asymptotic QNMs are independent of the value of 
$\g$, as long as it is finite. In contrast, if $\g=0$, we obtain $j_0=j_\infty=1$, and the determinant
condition \req{wkb-det} leads to $e^{i\nu x_0}=0$, which has no finite solution. This is in agreement
with the exact solution: there are no QNMs when $\g=0$ because the corresponding conductivity is
a constant function.

\end{document}